\documentclass[a4paper,11pt]{article}
\usepackage[margin=1in]{geometry}
\usepackage{epsfig}
\usepackage{cite}
\usepackage{graphicx}
\DeclareGraphicsExtensions{%
    .png,.PNG,%
    .pdf,.PDF,%
    .jpg,.mps,.jpeg,.jbig2,.jb2,.JPG,.JPEG,.JBIG2,.JB2}
\usepackage{epstopdf}
\usepackage{color}
\usepackage{tabularx}
\usepackage{amsmath}
\usepackage{comment}
\usepackage{amssymb}
\usepackage{mathtools}
\usepackage{float}
\usepackage{multirow}
\restylefloat{table}
\usepackage{booktabs}
\usepackage[dvipsnames]{xcolor}
\usepackage{lscape}
\date{\today}
\def\unit{\leavevmode\hbox{\small1\kern-3.6pt\normalsize1}}

\def\gtwid{\mathrel{\raise.3ex\hbox{$>$\kern-.75em\lower1ex\hbox{$\sim$}}}}
\def\ltwid{\mathrel{\raise.3ex\hbox{$<$\kern-.75em\lower1ex\hbox{$\sim$}}}}
\def\gev{{\rm \, Ge\kern-0.125em V}}
\def\tev{{\rm \, Te\kern-0.125em V}}

\def    \be            {\begin{equation}}
\def    \ee            {\end{equation}}
\def    \bea           {\begin{eqnarray}}
\def    \eea           {\end{eqnarray}}

\def\a{\alpha}
\def\b{\beta}

\def\d{\delta}
\def\n{\nu}

\def\m{\mu}
\def\nn{\nonumber}
\def\d{\delta}
\def\D{\Delta}
\def\s{\sigma}
\def\r{\rho}
\def\t{\theta}

 \normalsize

 \normalsize

\newcommand{\bmat}{\left(\begin{array}}
\newcommand{\emat}{\end{array}\right)}

\begin{document}
\renewcommand{\thefootnote}{\fnsymbol{footnote}}
\vspace{.3cm}
\title{\Large\bf {Role of unphysical neutrino phases in the phenomenology of consistently defined textures. Case study: traceless neutrino  mass  matrix}} \author
{  N. Chamoun$^{1,2,3}$\thanks{chamoun@uni-bonn.de}
and E. I. Lashin$^{4,5}$\thanks{slashin@zewailcity.edu.eg, elashin@ictp.it} ,
 \\\hspace{-0.cm}
 \footnotesize$^1$ Statistics Department, Faculty of Sciences, University of Damascus, Damascus, Syria. \\
 \footnotesize$^2$ CASP, Antioch Syrian University, Maaret Saidnaya, Damascus, Syria \\
  \footnotesize$^3$ Bethe Center for Theoretical Physics, Physikalisches Institut der Universit\"at Bonn, Nußallee 12, 53115 Bonn, Germany\\
\footnotesize$^4$  Department of Physics, Faculty of Science, Ain Shams University, Cairo 11566,  Egypt.  \\
\footnotesize$^5$ CFP, Zewail City of Science and
Technology, 
 \footnotesize  6 October City, Giza 12578, Egypt.
  \\\hspace{-0.cm}
 }
\date{\today}
\maketitle
\begin{abstract}
 We highlight the role played by the unphysical phases in the definition of neutrino mass matrix, showing it does not relate to mere semantics, rather it has an effect on the phenomenology. As a case study, we take the neutrino mass matrix texture characterized by a vanishing trace, and study the effect of the phases, physical and unphysical, in its definition. We undergo a thorough phenomenological analysis, first (second) when the unphysical (CP) phases are vanishing, then move on to the general case where all phases exist. We stress that the effect of the unphysical phases on the phenomenology originates from changing the texture definition upon introducing them, while they are not observable. Finally we present a theoretical realization of the texture based on $A_5$ non-abelian flavor symmetry, which is also related to a consistently defined texture.
\end{abstract}
\maketitle
{\bf Keywords}: Neutrino Physics; Flavor Symmetry;
\\
{\bf PACS numbers}: 14.60.Pq; 11.30.Hv;
\vskip 0.3cm \hrule \vskip 0.5cm'
\section{Introduction}
The examination of specific textures of Majorana neutrino mass matrix $M_\nu$ is
a traditional approach to the flavor structure in the lepton sector, and offers a window to physics beyond the standard model (SM). However,
the main motivation for this phenomenological approach is simplicity and
predictive power, which would be suggestive of some nontrivial
symmetries or other underlying dynamics. Many forms were studied, like the tri-bimaximal mixing (TBM)\cite{tbm}, the one zero element \cite{Rodejohan, 0texture}, the two zero elements \cite{Frampton_2002, Xing_2002}, the vanishing minor \cite{Lashin_2008}, the two vanishing subtraces \cite{Alhendi_2008, ismael_2022}, the two equalities \cite{2=_indian} and the hybrid of zero element and minor \cite{hybrid}.

In \cite{ismael_npb}, we studied the role played by the unphysical phases ($\phi^{\mbox{\tiny unphys}}$) in the definition of any texture. Although these unphysical phases may appear in some sectors beyond SM, however, after all symmetry breakings, they can be absorbed by the charged lepton fields and are thus non-physical in any setup involving just SM augmented by neutrino masses. Nonetheless, unphysical phases represent details of the mass matrix that are relevant for both processes of the mass matrix diagonalization and the renormaliztion group equations for physical parameters contained in the mass matrix \cite{Casas}. The authors of \cite{Adhikary} stressed the importance of the unphysical phases for the diagonalization, and by eliminating them could find relations involving only the physical parameters. However, they did not discuss the role played by the unphysical phases in the way a texture of a particular mathematical form is defined, a point we aim to address in this work.

 Being non physical, one might be tempted to expect that scanning over the unphysical phases should not lead to a phenomenology of the other physical parameters different from that when equating the unphysical phases to zero. This expectation does not hold for all textures, and that motivated us to study the link between the unphysical phases and the way one defines the texture in order to meet a given mathematical condition. We realized that forcing the unphysical phases to vanish corresponded to a different texture, albeit having the same mathematical form,  from when including them, and thus no reason for the phenomenology of the physical parameters to remain the same. 
  
In order to be consistent, the only way to make use of any study assuming vanishing unphysical phases, is to consider it as corresponding to a new texture definition, described as ``specific'', where the defining mathematical constraint ($g(M_\nu)=0$) is given in the vanishing unphysical phases slice. We mean by this that for $M_\n$ to meet the texture definition, one does not enforce ($g(M_\n)=0$), rather we force ($g(M^{\mbox{\tiny phys}}_\nu)=0$), where $M_\n^{\mbox{\tiny phys}}$ is equivalent to $M_\n$ but with vanishing $\phi^{\tiny unphys}$. In general, introducing the unphysical phases, would `dilute' the correlation plots, and no reason to assume them vanishing from the start if the objective was to study a particular texture defined by a mathematical condition and to determine its predictions. Moreover, any claim to realizing the texture, according to the ``specific'' definition, within a model will not suffice unless it is shown that it led to the requested texture while remaining at the same time in the vanishing unphysical phases slice. As the latter is parametrization dependent, then finding a realization model for the texture with such a ``specific'' definition represents a tough task. More precisely, we considered three definitions for a given texture, called ``mathematical'', ``specific'' and ``generalized'' (see Table \ref{properties}), where the first ``mathematical'' (second ``specific'') is sensitive to the unphysical phases (PMNS parametrization), whereas only the third ``generalized'' definition is `physically' sound. The correlations of the ``specific'' definition form a subset of those of the ``generalized'' definition, while, putting aside the $\phi^{\mbox{\tiny unphys}}$'s, the ``mathematical'' and ``generalized'' definitions have identical phenomenologies.

All these subtlties about unphysical phases were applied in \cite{chamoun2023} for a texture defined by two constraints (one equality and one anti-equality), whereas we apply them in this work to a texture characterized by zero trace ($\Sigma_{i=1}^{i=3}M_{\nu ii}=0$) --a condition which is not invariant under rephasing-- aiming to underscore the role of the unphysical phases in setting up the texture definition, and its effect on the phenomenological studies.   

In fact, the authors of \cite{nasri2000} considered first the vanishing trace condition, as it represented, in addition to the determinant, an invariant quantity upon carrying a similarity transformation, and a realization model based on $SO(10)$ was suggested. The traceless ansatz  was later furthered in \cite{nasri2004,nasri2005} with some applications to leptogenesis. One `zero' condition was taken in \cite{zee2003}, where the sum of the neutrino eigenmasses, up to `alternate' signs, was considered to be vanishing, which would give the traceless condition but only in the real case, and signs were introduced via the Majorana phases being equal to $0$ or $\pi/2$. In \cite{Rodejohann2004}, the traceless condition was retaken in both CP conservation/violation cases. However, in all these studies, the unphysical phases were assumed to be vanishing, and the resulting correlation plots were thus limited to a slice in the  parameter space characterized by vanishing unphysical phases.  Actually, recently, the question of traceless texture was retaken, albeit with no offer of any realization model, first in \cite{madan2018} and later in \cite{sangeeta2023}. However, these studies again were limited to the vanishing unphysical phases, and although some additional conjoint constraints were considered, however the traceless condition stated therein was rather a parametrization-dependent `sum rule' than a vanishing trace condition.

The objective of this paper is to update the traceless texture in view of the new data, and in particular to carry out a thorough analysis where all phases, be them CP or unphysical ones, are considered. While doing so, the role of the unphysical phases in the definition of the texture will be stressed and emphasized, contrasting various ways of defining the texture each with its own phenomenology. We shall consider three cases which can be thought of as three different textures with three different phenomenologies. We first (case I) switch off the unphysical phases, updating thus past studies, which should be looked upon in the context of the ``specific'' definition. Second, we examine the CP conserved situation (case II), where the physical phases are turned off but instead of just inserting signs at will in the zero mass-sum condition, we consider the full potential of the unphysical phases and see how they lead to many solutions diluting many correlation plots. Finally, we undergo a complete  analysis where the texture is bound only by its defining traceless mathematical condition (case III). 

The choice of the traceless texture to study the effect of the unphysical phases is quite suitable, as the three studied cases clarify well the intricate interplay between the mathematical traceless constraint and where to impose it. As mentioned above, (case I) parallels past studies, but we opted to do it in a different parametrization in order to show that the ``specific" definition is effectively parametrization dependent. The difference between cases I \& II resides in where to impose the traceless condition, but while the corresponding vanishing unphysical phases slice in (case I) is parametrization-dependent, the corresponding slice in (case II) of vanishing CP phases is physical and thus is independent of parametrization.  The third case corresponds to ``Generalized" definition which is insensitive both to  the unphysical phases choice and to the parametrization. We shall see also that the number of the free parameters in the three cases is different showing well we are concerned here with the study of three different textures, albeit defined by one mathematical constraint corresponding to vanishing trace. For the phenomenology in (case III), one can adopt either one of the  ``generalized'' or ``mathematical'' definitions, since they are equivalent as far as the `physical' correlations are concerned, whereas non-trivial correlations involving $\phi^{\mbox{\tiny unphys}}$ result only in the context of the ``mathematical'' definition. Moreover, we succeeded in finding a realization model, valid for both ``mathematical'' and ``generalized'' definitions, based on an $A_5$ non-abelian flavor symmetry, within seesaw type II scenario, but at the expense of enriching the matter content, where new scalars upon breaking spontaneously the flavor symmetry lead to the desired form of a traceless texture.


In summary, introducing the unphysical phases, although they are devoid of any physical meaning, in a physically plausible way, does change the definition of the texture under study, and so it is normal, putting aside the updating of the experimental data and comparing to case I (traceless texture according to ``specific'' definition), that one gets new phenomenologies for cases II (zero trace in the vanishing CP phases slice) and III (traceless texture according to ``generalized'' definition), since the three cases correspond to three different definitions of the traceless texture, whence three phenomenologies. However, the realization model presented corresponds to case III, whereas in past studies, assuming vanishing unphysical phases in a certain parametrization, the realization models presented were not complete as there was no check whether or not they corresponded to vanishing unphysical phases, which furthermore would depend on the chosen PMNS parametrization. 

The plan of the paper is as follows. We present the notations in section 2, give the various texture definitions and their characteristics in section 3, and define the traceless texture in section 4. In section 5, we carry out the phenomenological analysis starting in subsection 5.1 (5.2) with the case of vanishing unphysical (CP) phases, then treat the general case in subsection 5.3. In subsection 5.4, we comment on the number of free parameters accounting for the studied cases, giving in passing a ``good" texture definition valid for the general case illustrating its free parameters counting. We present in detail an $A_5$-realization for the texture in section 6, and we end up with conclusion and summary in section 7. Some technical details are given in appendices.

\section{Notations}
Working in the `flavor' basis, where the charged lepton mass matrix is diagonal, then the neutrino sector is wholly responsible for the observed neutrino mixing:
\begin{equation} \label{eq1}
V^{\dagger}M_{\nu}V^{*}= M_\n^{\mbox{\tiny diag.}} \equiv \mbox{diag}\left(m_1,m_2,m_3\right),
\end{equation}
with ($m_{i}, i=1,2,3$) real positive neutrino masses. The lepton mixing matrix $V$ contains three mixing angles, three CP-violating phases and three unphysical phases. We write it , in our adopted parameterization where the third column of $U_{\mbox{\tiny PMNS}}$ is real, as a Dirac mixing matrix $U_\d$ (consisting of three mixing angles and a Dirac phase) pre(post)-multiplied with a diagonal matrix $P_\phi$ ($P^{\mbox{\tiny Maj.}}$) consisting of three (two Majorana) phases:
\bea
\label{defOfU}
V & = & P_\phi\;U_\d\;P^{\mbox{\tiny Maj.}} \;: \; U_\d \; = R_{23}\left(\t_{23}\right)\; R_{13}\left(\t_{13}\right)\; \mbox{diag}\left(1,e^{-i\d},1\right)\; R_{12}\left(\t_{12}\right), \nn
\\
P_\phi &=& \mbox{diag}\left(e^{i\phi_1},e^{i\phi_2},e^{i\phi_3}\right)\;,\;P^{\mbox{\tiny Maj.}} = \mbox{diag}\left(e^{i\rho},e^{i\sigma},1\right),\nn \\
 U_{\mbox{\tiny PMNS}}& =& U_\d\;P^{\mbox{\tiny Maj.}} =   \left ( \begin{array}{ccc} c_{12}\, c_{13} e^{i\rho} & s_{12}\, c_{13} e^{i\sigma}& s_{13} \\ (- c_{12}\, s_{23}
\,s_{13} - s_{12}\, c_{23}\, e^{-i\delta}) e^{i\rho} & (- s_{12}\, s_{23}\, s_{13} + c_{12}\, c_{23}\, e^{-i\delta})e^{i\sigma}
& s_{23}\, c_{13}\, \\ (- c_{12}\, c_{23}\, s_{13} + s_{12}\, s_{23}\, e^{-i\delta})e^{i\rho} & (- s_{12}\, c_{23}\, s_{13}
- c_{12}\, s_{23}\, e^{-i\delta})e^{i\sigma} & c_{23}\, c_{13} \end{array}  \right ),\nn\\
\eea
where $R_{ij}(\theta_{ij})$ is the rotation matrix through the mixing angle $\theta_{ij}$ in the ($i,j$)-plane, ($\delta,\rho,\sigma$) are three CP-violating phases, and we denote ($c_{12}\equiv \cos\theta_{12}, s_{12}\equiv \sin\theta_{12}, t_{12}\equiv \tan\theta_{12}...)$.

For the sake of comparison, we provide the mixing matrix in the PDG parameterization as \cite{Rodejohann2004,fogli_2006}:
\bea
\label{defPDG}
V^{\mbox{\tiny PDG}} & =& P_\phi^{\mbox{\tiny PDG}}\; U_\d^{\mbox{\tiny PDG}}\; P^{\mbox{\tiny Maj.}, {\mbox{\tiny PDG}}}\;:\;
P_\phi^{\mbox{\tiny PDG}}=\mbox{diag}\left(e^{i\phi'_1},e^{i\phi'_2},e^{i\phi'_3}\right), P^{\mbox{\tiny Maj.}, {\mbox{\tiny PDG}}} = \mbox{diag}\left(1,e^{i\,\alpha},e^{i\,(\beta + \delta)}\right),\nn\\
 U_\delta^{\mbox{\tiny PDG}} &=&   \left ( \begin{array}{ccc} 
c_{12}\, c_{13}  & s_{12}\, c_{13} & s_{13}\,e^{-i\,\delta} \\ 
- c_{12}\, s_{23}
\,s_{13} \, e^{i\,\delta}- s_{12}\, c_{23} & - s_{12}\, s_{23}\, s_{13} \, e^{i\,\delta}+ c_{12}\, c_{23}
& s_{23}\, c_{13}\, \\
 - c_{12}\, c_{23}\, s_{13}\, e^{i\,\delta} + s_{12}\, s_{23} & - s_{12}\, c_{23}\, s_{13}\,e^{i\,\delta}
- c_{12}\, s_{23} & c_{23}\, c_{13} \end{array}   \right ).
\eea
 Imposing the invariance of $M_\n$ elements, upon using the two parameterizations in Eqs.(\ref{defOfU}, \ref{defPDG}),  amounts to keeping 
($\theta_{12}, \theta_{23}, \theta_{13}, \d$) the same, while having the transformation rules for the Majorana and unphysical phases depicted as follows.
\bea
\mbox{Adopted parametrization: }&& \left(\phi_1,\phi_2, \phi_3,  \r, \s \right),\;\;\;\mbox{PDG parametrization: }\;\;  \left(\phi'_1,\phi'_2, \phi'_3,  \alpha, \beta \right),\nn\\
\mbox{Transformation laws: } && \left(\phi_1 = \phi'_1 + \beta ,\; \phi_2=\phi'_2 + \beta + \d  , \; \phi_3=\phi'_3 + \beta +\d,\; \rho=- \beta,\;\sigma=\alpha - \beta \right).\nn\\
 \label{param}
\eea
 The rules of Eq. (\ref{param}) will allow us to contrast the numerical results of imposing vanishing unphysical phases in both parameterizations. Our ``Adopted" parametrization has the advantage of not showing the Dirac phase $\d$ in the effective mass term of the double beta decay \cite{xing_2001,xing_2002}. However, one should note the slice of vanishing unphysical phases in this parametrization does not correspond to `constant', let alone vanishing,  unphysical phases in other parametrizations as evident from the transformation laws in Eq.(\ref{param}).

The neutrino mass spectrum is divided into two classes: Normal hierarchy ($\textbf{NH}$) where $m_{1}<m_{2}<m_{3}$, and Inverted hierarchy ($\textbf{IH}$) where $m_{3}<m_{1}<m_{2}$. The solar and atmospheric neutrino mass-squared differences, and their ratio $R_{\nu}$, are given by:
\begin{equation}
\delta m^{2}\equiv m_{2}^{2}-m_{1}^{2},~~\Delta m^{2}\equiv\Big| m_{3}^{2}-\frac{1}{2}(m_{1}^{2}+m_{2}^{2})\Big|,     \;\;
R_{\nu}\equiv\frac{\delta m^{2}}{\Delta m^{2}}.\label{Deltadiff}
\end{equation}
with experimental data indicating ($R_{\nu}\approx 10^{-2}\ll1$). There are also two mass parameters measured experimentally, the
effective electron-neutrino mass:
\begin{equation}
\langle
m\rangle_e \; = \; \sqrt{\sum_{i=1}^{3} \displaystyle \left (
|V_{e i}|^2 m^2_i \right )} \;\; ,
\end{equation}
and the effective Majorana mass term
$\langle m \rangle_{ee} $:
\begin{equation} \label{mee}
\langle m \rangle_{ee} \; = \; \left | m_1
V^2_{e1} + m_2 V^2_{e2} + m_3 V^2_{e3} \right | \; = \; \left | M_{\n 11} \right |.
\end{equation}

The Jarlskog rephasing invariant quantity is given by \begin{equation}\label{jg}
J = s_{12}\,c_{12}\,s_{23}\, c_{23}\, s_{13}\,c_{13}^2 \sin{\delta}
\end{equation} and its value being non-vanishing is a necessary requirement for the presence of CP violation.  

Cosmological observations put bounds on the  `sum'
parameter $\Sigma$:
\be
\Sigma = \sum_{i=1}^{3} m_i.
\ee

The allowed experimental ranges of the neutrino oscillation parameters at  3$\sigma$ level with the best fit values are listed in Table(\ref{TableLisi:as}) \cite{de_Salas_2021}.
\begin{table}[h]
\centering
\scalebox{0.8}{
\begin{tabular}{cccc}
\toprule
Parameter & Hierarchy & Best fit &  $3 \sigma$ \\
\toprule
$\delta m^{2}$ $(10^{-5} \text{eV}^{2})$ & NH, IH & 7.50 &  [6.94,8.14] \\
\midrule
 \multirow{2}{*}{$\Delta m^{2}$ $(10^{-3} \text{eV}^{2})$} & NH & 2.51 &  [2.43,2.59] \\
 \cmidrule{2-4}
           & IH & 2.48 &  [2.40,2.57]\\
\midrule
$\theta_{12}$ ($^{\circ}$) & NH, IH & 34.30 &  [31.40,37.40] \\
\midrule
\multirow{2}{*}{$\theta_{13}$ ($^{\circ}$)}  & NH & 8.53 & [8.13,8.92] \\
\cmidrule{2-4}
    & IH & 8.58 &  [8.17,8.96]\\
\midrule
\multirow{2}{*}{$\theta_{23}$ ($^{\circ}$)}  & NH & 49.26 &  [41.20,51.33] \\
\cmidrule{2-4}
      & IH & 49.46 &  [41.16,51.25]   \\
\midrule
\multirow{2}{*}{$\delta$ ($^{\circ}$)}  & NH & 194.00 &  [128.00,359.00] \\
\cmidrule{2-4}
 & IH & 284.00 & [200.00,353.00]   \\
\bottomrule
\end{tabular}}
\caption{\footnotesize The experimental bounds for the oscillation parameters at 3$\sigma$-level, taken from the global fit to neutrino oscillation data \cite{de_Salas_2021} (the numerical values of $\Delta m^2$ are different from those in the reference which uses the definition $\Delta m^2 = \Big| \ m_3^2 - m_1^2 \Big|$ instead of Eq. \ref{Deltadiff}). Normal and Inverted Hierarchies are
respectively denoted by NH and IH}.
\label{TableLisi:as}
\end{table}

For the non-oscillation parameters, we adopt the 90\% C.L. upper limits, which are obtained by KATRIN and Gerda experiment for $m_{e}$ and $m_{ee}$ \cite{Aker_2019,Agostini_2019} . However, we adopt for $\Sigma$ the results of Planck 2018 \cite{Planck} from temperature information with low energy by using the simulator SimLOW.
\begin{equation}\label{non-osc-cons}
\begin{aligned}
\Sigma~~~~~&<0.54~\textrm{ eV},\\
 m_{ee}~~&<0.228~\textrm{ eV},\\
 m_{e}~~~&<1.1~\textrm{ eV}.
\end{aligned}
\end{equation}
We did not take the tight bound ($\Sigma<0.09~\textrm{ eV}$) \cite{Valentino_2021} using data from Supernovae Ia luminosity distances, neither the strict constraint of Planck 2018 combining baryon acoustic oscillation data in $\Lambda CDM$ cosmology ($\Sigma<0.12~\textrm{ eV}$) \cite{Planck}, nor the PDG live bound ($\Sigma<0.2~\textrm{ eV}$) \footnote{https://pdglive.lbl.gov/DataBlock.action?node=S066MNS} originating from fits assuming various cosmological considerations. Essentially, we opted for a more relaxed constraint for $\Sigma$ because we would like to give more weight to colliders' data compared to cosmological considerations in testing our particle physics model, and also because the tight bounds make use of cosmological assumptions which are far from being anonymous \cite{Chacko_2020}.

For simplification/clarity purposes regarding the analytical expressions, we from now on denote the mixing angles as follows.
\begin{equation}
\theta_{12}\equiv\theta_{x},~~\theta_{23}\equiv\theta_{y},~~\theta_{13}\equiv\theta_{z}.\label{redefine}
\end{equation}
However, we shall keep the standard nomenclature in the tables and figures for consultation purposes.

\section{How to define a texture?}

{\bf Three `` different" definitions of the texture}
\begin{itemize}
\item 
 ``Rephasing'' the neutrino mass matrix\footnote{i.e. redefining the neutrino gauge fields by phasing them, which can be absorbed by rephasing the charged lepton fields, and thus is not physical per se.}  amounts to multiplying $M_\n$ from right and from left by the diagonal unphysical phase matrix $P_\phi$. 
For SM supplemented by neutrino masses, we thus have equivalent descriptions using the two sets (look at Eq. \ref{defOfU}) 
\bea
\left(\n_i, M_{ij}\right) &\equiv& \left(\n^\prime_i = e^{-i\phi_i} \n_i, M^\prime_{ij}=e^{i(\phi_i  + \phi_j)} M_{\n ij} (\mbox{no sum}) \right)
\eea
 such that the two equivalent matrices are related by rephasing 
  \bea M^\prime_\n = P_\phi M_\n P_\phi &\sim& M_\n \eea
  
 Moreover, taking the `observable' $U_{\mbox{\tiny PMNS}}$ in a certain form or parametrization, imposes a corresponding parametrization for $P_\phi$ through\bea \label{parameterization}  V = P_\phi^{\mbox{\tiny Param}} U_{\mbox{\tiny PMNS}}^{\mbox{\tiny Param}}\eea

\item Texture definition to be consistent should be carried out at the level of the set of equivalence classes $\mathcal{M}/\sim$, where $\mathcal{M}$ is the set of $3\times3$ symmetric complex matrices, so that if one representative matrix meets the texture definition then all the equivalent matrices should also meet it.
     
\item For a texture given by a constraint of the form $g(M)=0$, where $g$ is a vector-valued function, we have the following different definitions:

\begin{enumerate}
\item  ``Mathematical" definition: \bea M \in \mbox{texture} &\Leftrightarrow& g(M)=0 \label{mathematical}\eea This definition is manifestly parametrization-independent but it is not rephasing invariant. In order to find all the matrices satisfying it we need to scan all the parameters including the unphysical phases.  

\item ``Specific" definition related to a ``slice" $S \subset \mathcal{M}$ \bea \label{specific} M \in \mbox{texture} &\Leftrightarrow& \exists M^\prime  \in S :  g(M^\prime)=0 \wedge M^\prime \sim M \eea This definition is manifestly rephasing invarinat, but in general can be parametrization-dependent if $S$ is. In order to find the matrices satisfying it, we can scan over parameters such that to remain in the slice $S$. 

\item ``Generalized" definition \bea \label{generalized} M \in \mbox{texture} &\Leftrightarrow& \exists M^\prime  : g(M^\prime)=0  \wedge M^\prime \sim M \eea  This definition is parametrization-independent and rephasing invariant.  One can look at it also as Specific definition related to the ``mathematical definition" slice $S^\prime=\{M\in \mathcal{M} : g(M)=0\}$. For this definition, it is desirable to find a ``good" function $g^{\mbox{\tiny good}}$ where any dependence on the unphysical phases drops out automatically such that  \bea \label{good} M \in \mbox{texture} &\Leftrightarrow \exists M^\prime  : g(M^\prime)=0  \wedge M^\prime \sim M \Leftrightarrow&  g^{\mbox{\tiny good}}(M)=0  \eea  However, it is not always possible to find such a ``good" defining function. 
\end{enumerate}

\end{itemize}

{\bf Requirements for a sound texture definition}
\begin{enumerate}
\item As mentioned before, a first requirement for a consistent texture definition, in order to be physically sound, is that it should be rephasing invariant, so that the two physically equivalent matrices $M_\n, M'_\n$ either both meet the texture definition or neither does.
    
\item A second requirement for a consistent texture definition is to be parameterization independent. We see now that any definition making use of a parameterization-dependent slice does not meet this requirement. Actually, changing the PMNS parametrization  may lead to changing the corresponding  Majorana  and unphysical phases, as we saw in Eq. (\ref{param}), where we see explicitly that a vanishing unphysical phases slice in one parametrization may not be so in another parametrization.   
 
\item A third important point concerns the realizability of the texture definition in model building. Thus, as was the case in many past studies, if the texture was defined according to the ``specific definition'' by a mathematical constraint on the $M_\n$-elements assuming vanishing unphysical phases, then it is not enough to conceive a model leading to the mathematical constraint, but rather it should lead to this constraint while at the same time assuring that the corresponding unphysical phases, within the taken parametrization, are vanishing as well.
\end{enumerate}

Table (\ref{properties}) summarizes the above properties for the various definitions. Moreover, the number of free parameters differs from definition to another in general, and depends usually on the specific slice used in the definition, if any. 

\begin{itemize}
\item
The mathematical definition (first line), as it is defined at the level of $M_\n$ elements, is parametrization independent, and once we find a model leading to the desired texture definition ($g(M_\n)=0$) then one can claim realizability. However it is not rephasing invariant in general \footnote{To clarify this last point, 
we start by a simple texture example, which is a one zero texture defined, say, by ($M_{\n11}=0$). We find that this texture definition is insensitive to the unphysical phases as we have also ($M'_{\n11}=0$), and this zero-texture is indeed rephasing-invariant.
However, if we take a texture definition given by ($M_{\n11}=M_{\n22}$), we see that ($M'_{\n11}=e^{2i\phi_1}M_{\n11}, M'_{\n22}=e^{2i\phi_2}M_{\n22}$), and so we get ($M'_{\n11} = e^{2i(\phi_1-\phi_2)}M'_{\n22} \neq M'_{\n22}$). Thus,  the texture definition is met for $M_\n$ whereas it is not met for  $M'_\n$.}.  To obtain the solutions according to this definition, one needs to scan over the full parameter space including the unphysical phases. 
\item
As said before, one should look at many past texture studies, assuming a mathematical constraint of the form ($g(M_\n)=0$) and restricting the analysis to vanishing unphysical phases, as being defined according to the ``specific'' definition (second line), with $S$ being the slice of vanishing unphysical phases. In practice, to get the solutions, i.e. the points in the parameter space satisfying the texture definition and meeting the experimental constraints, in all the 12-dim $M_\n$ space, one finds the solutions in the 9-dim  slice of vanishing unphysical phases, and then any obtained mass-matrix-solution would generate 3-dim many more phenomenologically equivalent mass matrices, differing only in unphysical phases, by just rephasing this obtained solution. Restricting, plausibly, the phenomenological analysis to physical parameters correlation plots, though, means that one can take the plots solely from the solutions in that vanishing slice. 

There are two drawbacks for this way of defining the texture. First, the definition is not parametrization independent (a vanishing unphysical phases slice in ``PDG'' parametrization would not correspond to vanishing  slice in our ``Adopted'' parametrization, and vice versa), so the definition starts by fixing  a parametrization, then  imposing the texture mathematical definition in the slice of vanishing unphysical phases according to the chosen parametrization. This is the way which was done in many past studies, and so the corresponding textures per se depended on the chosen PMNS parametrization. The second drawback is that none of the past studies, which introduced models leading to the desired form of the texture, checked that with the form they obtained, the unphysical phases were vanishing, and thus, in our opinion, the realization models they presented were not complete.
\item
The ``generalized" definition (third line) is both rephasing invariant and parametrization independent and does not need any additional checking once one finds a realization method to impose the mathematical constraint defining the texture. In practice, it corresponds to the mathematical definition augmented by rephasing, and so, restricting to the physical correlation plots, one can obtain its solutions, like the mathematical definition, by scanning over the unphysical phases as well as the other parameters.  
\end{itemize}

\begin{table}[h]
\centering
\scalebox{0.8}{
\hspace{-3cm}
\begin{tabular}{c||c|c|c|c|c|}
\toprule
$M_\n \in$ texture $\Leftrightarrow$ &$\phi^{\mbox{\tiny unphys.}}$-invariance &  Parametrization independence &  Physicality & $\phi^{\mbox{\tiny unphys}}$ correlations & realizability \\
\toprule \hline
 $g(M_\n)=0$ (`Mathematical' def.) & $\times$ & $\surd$ &  $\times$ & not trivial & $\surd$\\
 $g(M^{\mbox{\tiny{phys}}}_\n)=0$ (`Specific' def.) & $\surd$ & $\times$ &  $\times$ & trivial & $\times$\\
 $\exists M_\n' \sim M_\n: g(M_\n')=0$ (`Generalized' def.) & $\surd$ & $\surd$ &  $\surd$ & trivial & $\surd$ \\
\bottomrule
\end{tabular}}
\caption{\footnotesize Properties of the three different texture definitions. $\phi^{\mbox{\tiny unphys}}$-invariance means that the definition is defined for the equivalence class of matrices, where $M_\n' \sim M_\n$ means that both matrices have the same 9 physical observables and where $M_\n^{\mbox{\tiny{phys}}}$ is equivalent to $M_\n$ but with vanishing $\phi^{\mbox{\tiny unphys}}$. Because $\phi^{\mbox{\tiny unphys}}$'s are sensitive to the PMNS paramterization, then `Physicality' requires both $\phi^{\mbox{\tiny unphys}}$-invariance and PMNS parametrization-independence. By realizability we mean whether the model leading to a texture of the specified form can embody or not the definition.}
\label{properties}
\end{table}


\section{Traceless Texture}
Before we delve into the numerical analysis of the traceless texture, we stress the importance of the included phases. By switching off all the CP and unphysical phases, then $V$ in Eqs.~(\ref{eq1},\ref{defOfU},\ref{defPDG}) would be real and orthogonal, and as a consequence one would get for the traceless texture the condition ($\Sigma_i m_i =0$) which is absurd in view of the masses being positive and not all equal to zero simultaneously. Thus, as was mentioned in the introduction section, we shall consider three cases, starting with switching off the unphysical phases (case I), which could then be compared to previous studies of the taceless texture and which should be looked at as corresponding to a special slice of the parameter space. Then, we shall consider the CP invariant case (case II), where we switch off the Dirac and Majorana phases $(\d, \r, \s)$, but restore the unphysical phases, and last we study numerically the general case (case III) where both unphysical and CP phases are present.

We can precise which definition we took in the manuscript as follows. 
\begin{itemize}
\item case I: ``Specific" Definition with $S$ being the slice of vanishing unphysical phases. This slice is parametrization-dependent. We did it in ``Adopted" parametrization, whereas past studies did it in ``PDG" parametrization.  When scanning we forced the unphysical phases to be zero. 
\item case II: we applied ``Specific" definition with $S$ being the slice of CP conservation. This slice is parametrization-independent. When scanning, we forced the CP phases to be zero. 
\item case III: we applied ``Generalized" definition.  We scanned over all parameters. 
\item for completeness, we plotted in case III (II) correlations involving unphysical phases. However, these correlations are not physical, and they involve the unphysical phases found when scanning in the vanishing trace slice $S^\prime$ (and that of CP conservation).   
\end{itemize}

Using Eqs. (\ref{eq1}, \ref{defOfU}, \ref{defPDG}), we have
\begin{align}
M_{\nu~ab} = \Sigma_i V_{ai} V_{bi} m_i  \Rightarrow \mbox{Tr} (M_{\n})= \Sigma_i A_i m_i : A_i = \Sigma_a (V_{ai})^2 \label{Traceconds}
\end{align}
Table \ref{AsAnalyticExpressions} shows the analytical expressions of the A's in terms of the mixing, CP phase and unphysical phase angles for both the Adopted and the PDG parameterizations.
Note however that $e^{2i \r}$ ($e^{2i\s}$) being a common factor in $A_1$ ($A_2$) for the Adopted parameterization, and the same for  $e^{2i \alpha}$ ($e^{2i\beta}$) in $A_2$ ($A_3$) for the PDG parameterization, are  consequences of the particular form of parameterizations we took in Eqs. (\ref{defOfU}, \ref{defPDG}).

\begin{table}[h]
\begin{center}
\begin{tabular} {c||c}
\hline
\hline
Case I & vanishing unphysical phases \\
\hline
$A_1$ Adopted  & $e^{2i \r} (c_x^2 + s_x^2 e^{-2i\d})$ \\
$A_1$ PDG  & $c_x^2 \, s_z^2 e^{2i\d} + s_x^2  + c_x^2\, c_z^2$ \\
$A_2$  Adopted  & $e^{2i \s} (s_x^2 + c_x^2 e^{-2i\d})$ \\
$A_2$  PDG  & $e^{2i \alpha} \left[s_x^2\,s_z^2\  +  e^{2i\d}\, \left(c_z^2 + c_x^2\,s_z^2\right)\right]$ \\
$A_3$ Adopted  & $1$ \\
$A_3$  PDG  & $e^{2i \beta} \left(s_z^2\  + e^{2i\d}\,c_z^2 \right)$ \\
\hline
\hline
Case II & vanishing CP phases \\
\hline
$A_1$ Adopted  & $ e^{2i\phi_1} c_x^2 c_z^2 + e^{2i\phi_2} (c_x s_y s_z + s_x c_y)^2 + e^{2i\phi_3} (-c_x c_y s_z +s_x s_y)^2$ \\
$A_1$ PDG  & $ e^{2i\phi'_1} c_x^2 c_z^2 + e^{2i\phi'_2} (c_x s_y s_z + s_x c_y)^2 + e^{2i\phi'_3} (-c_x c_y s_z +s_x s_y)^2$ \\
$A_2$  Adopted  & $  e^{2i\phi_1} s_x^2 c_z^2 + e^{2i\phi_2} (-s_x s_y s_z + c_x c_y )^2  + e^{2i\phi_3} (s_x c_y s_z +c_x s_y )^2$ \\
$A_2$  PDG  & $  e^{2i\phi'_1} s_x^2 c_z^2 + e^{2i\phi'_2} (-s_x s_y s_z + c_x c_y )^2  + e^{2i\phi'_3} (s_x c_y s_z +c_x s_y )^2$ \\
$A_3$   Adopted & $e^{2i \phi_1}s_z^2 + e^{2i \phi_2}s_y^2 c_z^2 + e^{2i \phi_3}c_y^2 c_z^2$ \\
$A_3$   PDG & $e^{2i \phi'_1}s_z^2 + e^{2i \phi'_2}s_y^2 c_z^2 + e^{2i \phi'_3}c_y^2 c_z^2$ \\
\hline
\hline
Case III & general \\
\hline
$A_1$  Adopted  & $e^{2i \r} \left[ e^{2i\phi_1} c_x^2 c_z^2 + e^{2i\phi_2} (c_x s_y s_z + s_x c_y e^{-i \d})^2 + e^{2i\phi_3} (c_x c_y s_z -s_x s_y e^{-i\d})^2\right]$ \\
$A_1$  PDG  & $e^{2i\phi'_1} c_x^2 c_z^2 + e^{2i\phi'_2} (c_x s_y s_z\,e^{i\,\d} + s_x c_y )^2 + e^{2i\phi'_3} (c_x c_y s_z\, e^{i\,\d} -s_x s_y )^2$ \\
$A_2$ Adopted  & $e^{2i \s} \left[ e^{2i\phi_1} s_x^2 c_z^2 + e^{2i\phi_2} (-s_x s_y s_z + c_x c_y e^{-i \d})^2  + e^{2i\phi_3} (s_x c_y s_z +c_x s_y e^{-i\d})^2\right]$ \\
$A_2$ PDG  & $e^{2i \alpha} \left[ e^{2i\phi'_1} s_x^2 c_z^2 + e^{2i\phi'_2} (-s_x s_y s_z\,e^{i\,\d} + c_x c_y)^2  + e^{2i\phi'_3} (s_x c_y s_z\,e^{i\,\d} +c_x s_y )^2\right]$ \\
$A_3$ Adopted  & $e^{2i \phi_1}s_z^2 + e^{2i \phi_2}s_y^2 c_z^2 + e^{2i \phi_3}c_y^2 c_z^2$ \\
$A_3$ PDG  & $e^{2\,i\,\beta}\left(e^{2i \phi'_1}s_z^2 + e^{2i \phi'_2}s_y^2 c_z^2 \,e^{2\,i\,\d}+ e^{2i \phi'_3}c_y^2 c_z^2\, e^{2\,i\,\d} \right)$ \\
\hline
\hline
\end{tabular}

 \end{center}
 \caption{The analytical expressions for the coefficients A's in the considered cases of the traceless texture for both Adopted and PDG parameterizations.}
\label{AsAnalyticExpressions}
\end{table}

Furthermore, the coefficients $(A_1, A_2, A_3)$ defining the tracelss texture and presented in Table~(\ref{AsAnalyticExpressions}) reveal that upon changing the parameterizations, according to the transformation laws in Eq.(\ref{param}), the coefficients keep the same functional dependence on the mixing and phase angles  for  the cases II and III while in case I the functional dependence does change. This leads to far reaching consequences on the resulting phenomenology which means getting different phenomological 
results in case I depending on which parameterization we have chosen, in contrast to the cases II and III where the same phenomenological consequence are obtained irrespective of the used parametetizations. These finding are confirmed by our numerical studies and at the same time indicating the importance of including the unphysical phases to get phenomology independent of any used parameterization. 

 The vanishing of the whole trace condition can be expressed as follows,
\begin{align} \label{conds}
\mbox{Re}(A_1) m_1 + \mbox{Re}(A_2) m_2 +  \mbox{Re}(A_3) m_3   &=0,\nonumber\\
\mbox{Im}(A_1) m_1 + \mbox{Im}(A_2) m_2 +  \mbox{Im}(A_3) m_3   &=0.
\end{align}
By writing Eqs. (\ref{conds}) in a matrix form, we obtain
\begin{equation}
\left( \begin {array}{cc} \mbox{Re}(A_1)&\mbox{Re}(A_2)\\ \noalign{\medskip}\mbox{Im}(A_1)&\mbox{Im}(A_2)\end {array}
 \right)\left( \begin {array}{c} m_{13}\\ \noalign{\medskip}m_{23}\end {array} \right)=-\left( \begin {array}{c} \mbox{Re}(A_3)\\ \noalign{\medskip}\mbox{Im}(A_3)\end {array} \right),
\end{equation}
where
\begin{align}
m_{13}=\frac{m_1}{m_3} &,& m_{23}=\frac{m_2}{m_3}
\end{align}
Solving Eqs. (\ref{conds}), we obtain
\begin{align}
m_{13}&=&\frac{\mbox{Re}(A_2)\mbox{Im}(A_3)-\mbox{Im}(A_2)\mbox{Re}(A_3)}{\mbox{Re}(A_1)\mbox{Im}(A_2)-\mbox{Re}(A_2)\mbox{Im}(A_1)},\nonumber\\
m_{23}&=&\frac{\mbox{Im}(A_1)\mbox{Re}(A_3)-\mbox{Re}(A_1)\mbox{Im}(A_3)}{\mbox{Re}(A_1)\mbox{Im}(A_2)-\mbox{Re}(A_2)\mbox{Im}(A_1)}.
\end{align}
Therefore in our adopted parameterization, we get the mass ratios in terms of the mixing angles ($\t=\t_x, \t_y, \t_z$), the CP phases ($\d, \r, \s$) and the unphysical phases ($\phi=\phi_1, \phi_2, \phi_3$).

The neutrino masses are written as
\begin{equation}
m_3=\sqrt{\frac{\delta m^2}{m_{23}^2-m_{13}^2}},~~m_1=m_3\times m_{13},~~m_2=m_3\times m_{23}.\label{spectrum}
\end{equation}
As we see, we have ten input parameters corresponding to $(\theta_{x},\theta_{y},\theta_{z},\delta,\r, \s, \delta m^2, \phi_1, \phi_2, \phi_3)$, which together with two real constraints in Eq.~(\ref{conds}) allows us to determine the twelve degrees of freedom in $M_{\nu}$.

\section{Phenomenological Analysis}

We scan 10 parameters in their allowed regions (with the measurable observables spanning their 3-sigma experimental ranges),  which, with the texture 2 real constraints, allow to reconstruct the 12-dim $M_\n$. We take these scanned parameters as being the angles (3 mixings and 3 CP phases and 3 unphysical phases) and the solar mass squared $\d m^2$. Any parameter space point thus gives a corresponding $M_\n$ which is tested whether it passes the atmospheric mass squared ($\D m^2$) 3-sigma constraint and those of Eq. \ref{non-osc-cons}. We could have taken ($\D m^2$) as the input mass scale but we preferred to take $\d m^2$ due to its narrower acceptable band. In this manner, our parameter space is 10-dim (in case III) or 7-dim (in case II or case I). Although the number of free parameters is large, however constraining them to be in their experimentally allowed regions would not make it trivial that the resulting $M_\n$ passes the remaining 4 constraints, which leads to limited acceptable points, whence possibility of clear correlations. 

We throw a large number $N$ of order $10^7-10^8$ of points randomly in the parameter space, and we get the corresponding acceptable points, from which we can get the 2D correlation plots. Note that one had to do the scanning/analysis for each hierarchy ordering since the experimental constraints are not the same in the two hierarchy types. 

In Table \ref{Predictions}, we summarize the numerical ranges predicted by the traceless texture in all three cases. We checked that by increasing the number of points in the parameter space, the correlations plots do not change much in form, but rather they become only denser. Thus, Table \ref{Predictions} of the texture acceptable ranges would not change much with increasing $N$.

 \newgeometry{left=3cm,bottom=0.05cm}
\begin{landscape}
\begin{table}[h]
\begin{center}
\begin{tabular} {c||c|c||c|c||c|c}
\hline
\hline
Observable & \multicolumn{2} {c}  {Case I: vanishing unphysical phases} &
\multicolumn{2}  {c} {Case II: vanishing CP phases} &
\multicolumn{2} {c} {Case III: general case}  \\
\hline
\hline
Hierarchy & {\bf I} & {\bf N} & {\bf I} & {\bf N} & {\bf I} &  {\bf N}  \\
\hline
$\theta_{12}\equiv \t_x$ ($^\circ$) & $31.40 - 37.40$ & $31.40-37.40$ & $31.40-37.40$ & $31.40-37.40$  & $31.40 - 37.40$ & $31.40-37.40$   \\
\hline
$\theta_{23} \equiv \t_y$ ($^\circ$) &$41.16 - 51.25$ & $41.20-51.33$ & $41.16 - 51.24$ & $41.20-51.32$ &$41.16 - 51.25$ & $41.20-51.33$\\
\hline
$\theta_{13}\equiv \t_z$ ($^\circ$) &$8.17-8.96$ & $8.13-8.92$ & $8.17-8.96$ & $8.13-8.92$  &$8.17-8.96$ & $8.13-8.92$\\
\hline
$\delta$ ($^\circ$) & $200.06-352.95$ & $128.05-240.50$ $\cup$ $293.50-359.00$ & $0$  & $0 $ & $200.22-353$ & $128.57-358.81$ \\
\hline
$\rho$ ($^\circ$) & $0.94-85.04 \cup 95.43-178.51$ & $53.06-125.82$ & $0$ & $0$ &$0.06-179.96$ & $0.04- 179.94$\\
\hline
$\sigma$ ($^\circ$) &$0.07 - 179.97 $ &$30.76- 147.11$ & $0$ & $0$ &$0.04- 179.99$ & $0.03- 179.88$ \\
\hline
$\phi_1$ ($^\circ$) & $0 $ & $0$ & $0.73 - 359.97$ & $0.03-359.42$ &$0.25-359.98$ & $0.75-359.71$ \\
\hline
$\phi_2$ ($^\circ$) &$0$ & $0$ & $0.91 - 358.53$ & $0.16-359.72$ & $0.12-359.98$ & $0.11-359.72$\\
\hline
$\phi_3$ ($^\circ$) &$0 $ & $0$ & $0.03-359.53$ & $0.16-359.88$ & $0.25-359.82$ & $0.01-359.82$\\
\hline
$m_1{\mbox{(eV)}}$ &$0.0486-0.1390$ & $0.0281-0.1740$ & $0.0486-0.1591$ & $7.86\times 10^{-7}-0.0793$  &$0.0486-0.1701$ & $2.56 \times 10^{-5}-0.1247$\\
\hline
$m_2{\mbox{(eV)}}$ &$0.0494-0.1393$ & $0.0293-0.1742$ & $0.0494-0.1593$ &$0.0083-0.0798$ &$0.0494-0.1704$ &$0.0083-0.1277$\\
\hline
$m_3{\mbox{(eV)}}$ &$0.0002-0.1299$ & $0.0572-0.1813$ & $1.81 \times 10^{-5}-0.1511$ & $0.0497-0.0942$ &$2.73\times 10^{-6}-0.1628$ &$0.0497-0.1371$\\
\hline
$m_e{\mbox{(eV)}}$ &$0.0483-0.1389$ & $0.0295-0.1742$ & $0.0483-0.1590$ & $0.0084-0.0798$ &$0.0483-0.1700$ & $0.0085-0.1278$\\
\hline
$m_{ee}{\mbox{(eV)}}$ &$0.0123-0.1338$ & $0.0264-0.1442$ & $0.0477-0.1590$ & $0.0033-0.0798$ &$0.0124-0.1623$ & $0.0002-0.0959$\\
\hline
$\Sigma {\mbox{(eV)}}$ & $0.0984-0.4082$ & $0.1147-0.5296$ & $0.0984-0.4696$ & $0.0583-0.2533$ & $0.0980-0.5033$ & $0.0582-0.3923$\\
\hline
\hline
 \end{tabular}
 \end{center}
 \caption{The various predictions for the ranges of the neutrino measurable parameters for the traceless texture in its three cases (vanishing unphysical phases, vanishing CP phases, general) at 3-$\sigma$ level.}
\label{Predictions}
 \end{table}
\end{landscape}
\restoregeometry

We introduce 15 correlation plots for  case I in either hierarchy ordering, generated from the accepted points of the neutrino physical parameters at the 3-$\sigma$ level. The first and second rows represent the correlations between the mixing angles and the CP-violating phases. The third row introduces the correlations amongst the CP-violating phases, whereas the fourth one represents the correlations between the Dirac phase $\delta$ and each of $J$, $m_{ee}$ and $m_2$ parameters respectively. The last row shows the degree of mass hierarchy plus the ($m_{ee}, m_2$) correlation. However, for cases II and III, and since the inclusion of the unphysical phases leads to the disappearance and dilution of many correlations, we kept just the apparent correlations, consisting of six ones divided into two rows. The first row represents the intercorrelations between ($\t_x, \t_y$ and $\phi_1-\phi_3$), whereas the second one lists the correlation ($\phi_2-\phi_3, \phi_1-\phi_3$) and the two mass correlations involving ($m_{ee}, m_2$) and ($m_{21}=m_2/m_1, m_3$).

Had the analytic expression of the zeros of ($m_{23}^2-m_{13}^2$) been easy to find, one could use these zeros as a good approximation offering an analytical justification, for the acceptable parameter space meeting $R_\n \approx 10^{-2}$ close to zero. However, that was not the case in our study, and we could only limit ourselves to the numerical computations.

Finally, we reconstruct $M_{\nu}$ for each viable ordering, at one representative point, chosen to be as close as possible to the best fit experimental values at the 3-$\sigma$ level. Note that the ranges of Table (\ref{Predictions}) represent projections of the acceptable points of the parameter space onto the corresponding axis of interest, and thus picking up projections may not correspond to one acceptable point. For reconstructing $M_\n$ we tried to get an acceptable point whose projections of $\t_{12}, \t_{23}$ and $\d$ are as close as possible to their central values, which may not correspond to these projections being equal simultaneously to their best-fit values even when these best-fit values fall in the acceptable regions.    

\subsection{Vanishing unphysical Phases}
The mass ratios are given as (``Num (Den)" stand for ``Numerator (Denominator))
\bea
\label{mass-ratios-zero-phi}
m_{13}=\frac{\mbox{Num}(m_{13})}{\mbox{Den}(m_{13})} &,& m_{23}=\frac{\mbox{Num}(m_{23})}{\mbox{Den}(m_{23})}: \\
\mbox{Num}(m_{13}) = c_x^2 s_{2\d-2\s}-s_x^2 s_{2\s} &,& \mbox{Num}(m_{23}) = s_x^2 s_{2\r-2\d}+c_x^2 s_{2\r}  \nn \\
\mbox{Den}(m_{13})= \mbox{Den}(m_{23}) &=& -s_x^2 c_x^2 s_{2\d-2\s}c_{2\r-2\d}+s_x^4 s_{2\s} c_{2\r-2\d}
+ (-s_x^2 c_x^2 s_{2\r-2\d} -s_{2\r}c_x^4)c_{2\d-2\s} \nn\\
&& -c_{2\s}s_x^4 s_{2\r-2\d}-c_x^4c_{2\r} s_{2\d-2\s} -s_x^2 c_x^2 s_{2\r-2\s}, \nn
\eea
whence one finds
\bea
\label{zerosm2-m1}
m_{23}^2-m_{13}^2 &=& \frac{\mbox{Num}(m_{23}^2-m_{13}^2 )}{\mbox{Den}(m_{23}^2-m_{13}^2 )}: \nn \\
\mbox{Num}(m_{23}^2-m_{13}^2 ) &=& \left[(s_{2\r} + s_{2\s} -s_{2\r-2\d}+s_{2\d-2\s})c_x^2 -s_{2\s}+s_{2\r-2\d} \right] \nn \\ && \times \left[ (s_{2\r} - s_{2\s} -s_{2\r-2\d}-s_{2\d-2\s})c_x^2 +s_{2\s}+s_{2\r-2\d} \right]
\eea

 Noting that no dependence on $\t_y$ and $\t_z$ for these zeros, one can a priori deduce that no correlations involving these two mixing angles exist. 

 From Table \ref{Predictions}, neither $m_1$ for normal hierarchy nor $m_3$ for inverted hierarchy does approach a vanishing value, so no singular such textures can accommodate the data.

\subsubsection{Inverted Hierarchy}
\begin{figure}[hbtp]
\hspace*{-4cm}
\includegraphics[width=24cm, height=16cm]{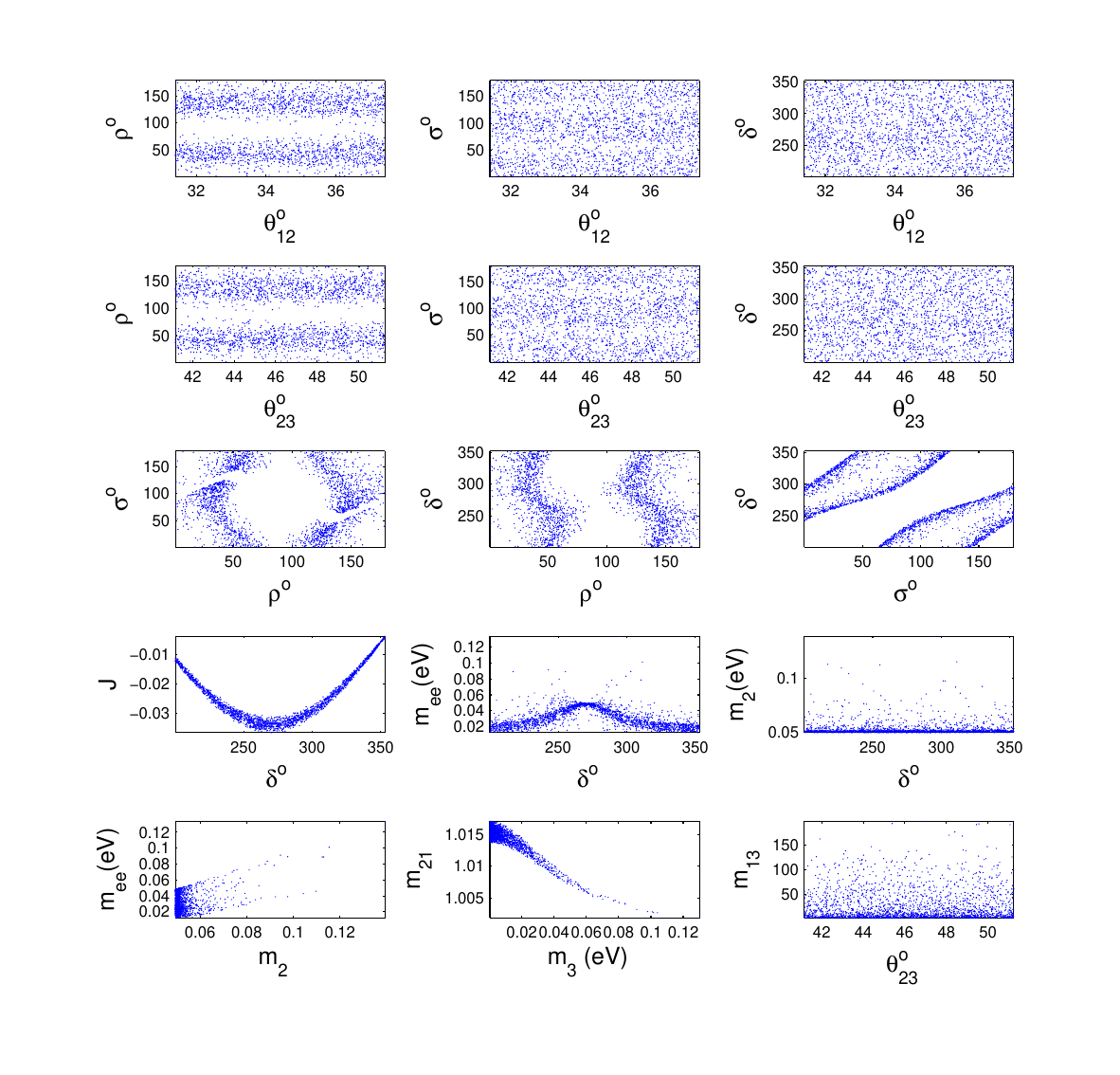}
\caption{The correlation plots for traceless texture in case I (vanishing unphysical phases), in the inverted ordering hierarchy. The first and second row represent the correlations between the mixing angles ($\theta_{12}$,$\theta_{23}$) and the CP-violating phases. The third row introduces the correlations amdist the CP-violating phases, whereas the fourth one represents the correlations between the Dirac phase $\delta$ and each of $J$, $m_{ee}$ and $m_2$ parameters respectively. The last row shows the degree of mass hierarchy plus the ($m_{ee}, m_2$) correlation.}
\label{caseI-inv}
\end{figure}

Fig. (\ref{caseI-inv}) shows the correlations corresponding to inverted hierarchy type.

As expected, no correlations involving $\t_y$ exist since the coefficients $A_i$'s do not depend on this parameter. We note equally that no clear correlations involve $\t_x$, whereas the correlations ($J,\d$) follows a usual sine curve ($J \propto \sin \d$) where the proportionality factor depends slightly on the mixing angles. We note also a small disallowed gap for $\r$ around $\pi/2$

Regarding the mass parameters, we notice a correlation ($m_{ee}, \d$), which originates from those amidst ($\r, \d$) and ($\s, \d$). The correlation ($m_{ee}, m_2$) shows that $m_{ee}$ increases with increasing $m_2$, and that the hierarchy can be mildly strong reaching ($m_{13} \approx 10^2$), so $m_3$ can not vanish.

We reconstruct the neutrino mass matrix for a representative point, which is taken such that the mixing angles and the Dirac phase are chosen to be as near as possible from their best fit values from Table \ref{TableLisi:as}. For inverted ordering, the representative point is taken as follows.
\begin{equation}
\begin{aligned}
(\theta_{12},\theta_{23},\theta_{13})=&(35.74^{\circ},48.58^{\circ},8.55^{\circ}),\\
(\delta,\rho,\sigma)=&(290.41^{\circ},28.58^{\circ},74.69^{\circ}),\\
(m_{1},m_{2},m_{3})=&(0.0491\textrm{ eV},0.0499\textrm{ eV},0.0027\textrm{ eV}),\\
(m_{e},m_{ee}, \Sigma)=&(0.0489\textrm{ eV},0.0352\textrm{ eV}, 0.1017 \textrm{ eV}),
\end{aligned}
\end{equation}
 and the corresponding neutrino mass matrix (in eV) is
\begin{equation}
M_{\nu}=\left( \begin {array}{ccc} 0.0029+0.0351i &  -0.0029 - 0.0259i &  0.0033 + 0.0214i\\ \noalign{\medskip}-0.0029 - 0.0259i &   -0.0003-0.0103i &  0.0041+0.0176i
\\ \noalign{\medskip} 0.0033 + 0.0214i &   0.0041+0.0176i &  -0.0026-0.0248i\end {array} \right).
\end{equation}

\subsubsection{Normal Hierarchy}
\begin{figure}[hbtp]
\hspace*{-4cm}
\includegraphics[width=24cm, height=16cm]{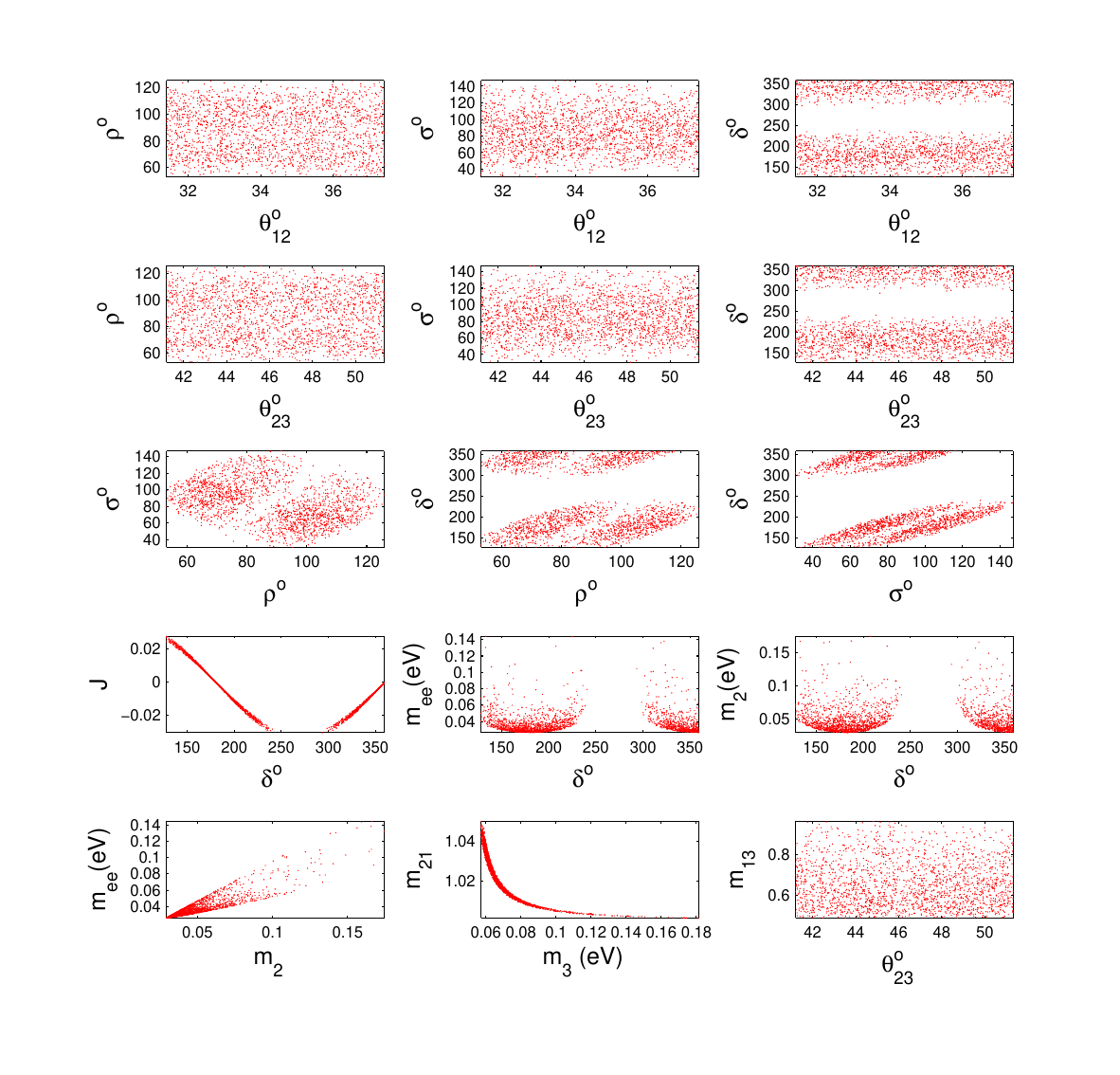}
\caption{The correlation plots for traceless texture in case I (vanishing unphysical phases), in the normal ordering hierarchy. The first and second row represent the correlations between the mixing angles ($\theta_{12}$,$\theta_{23}$) and the CP-violating phases. The third row introduces the correlations amdist the CP-violating phases, whereas the fourth one represents the correlations between the Dirac phase $\delta$ and each of $J$, $m_{ee}$ and $m_2$ parameters respectively. The last row shows the degree of mass hierarchy plus the ($m_{ee}, m_2$) correlation.}
\label{caseI-nor}
\end{figure}

Fig. (\ref{caseI-nor}) shows the correlations corresponding to normal hierarchy type.

Again, no correlations involving $\t_y$ as the coefficients $A_i$'s are $\t_y$-independent. Also,  no clear correlations involving $\t_x$, whereas the correlations ($J,\d$) is always sinusoidal in shape. 
There are disallowed gaps for ($\r, \s$ and $\d$). It is worthy to mention that upon using the PDG parameterization in our numerical study, we find disallowed gaps for $(\alpha , \beta)$ but not for $\delta$. 
This is not surprising since vanishing unphysical phases in one paraemetrization does not imply them to be vanishing in the other one, as can be checked from the transformation laws in Eq.~(\ref{param}).

Regarding the mass parameters, the correlation ($m_{ee}, \d$) results from those between ($\r, \d$) and ($\s, \d$). The correlation ($m_{ee}, m_2$) shows that $m_{ee}$ is increasing with respect to $m_2$, and that the hierarchy is weak characterized by ($m_1 \approx m_2 \approx m_3$), so $m_1$ can not vanish.

We reconstruct the neutrino mass matrix for a representative point, which is taken such that the mixing angles and the Dirac phase are chosen to be as near as possible from their best fit values from Table \ref{TableLisi:as}. For inverted ordering, the representative point is taken as follows.
\begin{equation}
\begin{aligned}
(\theta_{12},\theta_{23},\theta_{13})=&(35.81^{\circ},49.33^{\circ},8.56^{\circ}),\\
(\delta,\rho,\sigma)=&(192.02^{\circ},98.22^{\circ},93.97^{\circ}),\\
(m_{1},m_{2},m_{3})=&(0.0293\textrm{ eV},0.0305\textrm{ eV},0.0581\textrm{ eV}),\\
(m_{e},m_{ee}, \Sigma)=&(0.0306\textrm{ eV},0.0277\textrm{ eV}, 0.1179 \textrm{ eV}),
\end{aligned}
\end{equation}
and the corresponding neutrino mass matrix (in eV) is
\begin{equation}
M_{\nu}=\left( \begin {array}{ccc} -0.0269-0.0068i &  0.0101 - 0.0006i &  0.0079 + 0.0022i\\ \noalign{\medskip}0.0101 - 0.0006i &   0.0198+0.0032i &  0.0422-0.0036i
\\ \noalign{\medskip} 0.0079 + 0.0022i &   0.0422-0.0036i &  0.0071+0.0036i\end {array} \right).
\end{equation}

\subsection{Vanishing CP Phases}
In a similar way to case I, we impose the mathematical constraint (traceless) on the slice characterized by vanishing CP-phases. Thus a matrix belongs to the texture if and only if its equivalent matrix with vanishing CP-phases is traceless. While one would have be tempted, like the ``specific" definition in case I, to think that this definition is parametrization dependent, but the fact that eventhough the values of the CP-phases differ from parametrization to another in general, but the vanishing CP-phases slice is the slice corresponding to CP conservation, so it is physical and as such parametrization-independent. This means vanishing CP-phases in one parametrization leads to them vanishing in all parametrizations, as could be seen in Eq. \ref{param}: ($\r=\s=\d=0 \Leftrightarrow \a=\b=\d=0$)). Thus case II is both rephasing- and parametrization-independent.

We note from table (\ref{AsAnalyticExpressions}) that including the unphysical phases $\phi$ brings up a dependence on ($\t_x, \t_y$) to the $A_i$'s coefficients. The complex nature of $M_\n$ results from non-vanishing unphysical phases while CP-phases are vanishing.
Also, the numerical computations reveal a strong correlation between the unphysical phases, say ($\phi_1, \phi_2$). This comes because we have:
\bea
\label{massmatrixelementsphase}
{M_\n}_{ab} (\phi \neq 0) = {M_\n}_{ab} (\phi =0) e^{i (\phi_a + \phi_b)}
&\Rightarrow& {M_\n}_{kk} (\phi \neq 0) = e^{2i \phi_k} {M_\n}_{kk} (\phi =0)
\eea
and so the phases play an important role in meeting the traceless texture condition ($M_{\n 11} + M_{\n 22} + M_{\n 33} =0 $), since, as mentioned earlier, this condition can not be met with no phases at all. Factoring out $e^{2i \phi_3}$, we drew the correlation ($\phi_1-\phi_3, \phi_2-\phi_3 $).

We noted that although $m_1$ ($m_3$) in the normal (inverted) hierarchy type can reach tiny values of order of ${\cal O}(10^{-7})\left({\cal O}(10^{-5})\right)$, however the singular texture with these masses equaling exactly zero was not achieved and singular textures are not accommodated.

\subsubsection{Inverted Hierarchy}
\begin{figure}[hbtp]
\hspace*{-4cm}
\includegraphics[width=24cm, height=16cm]{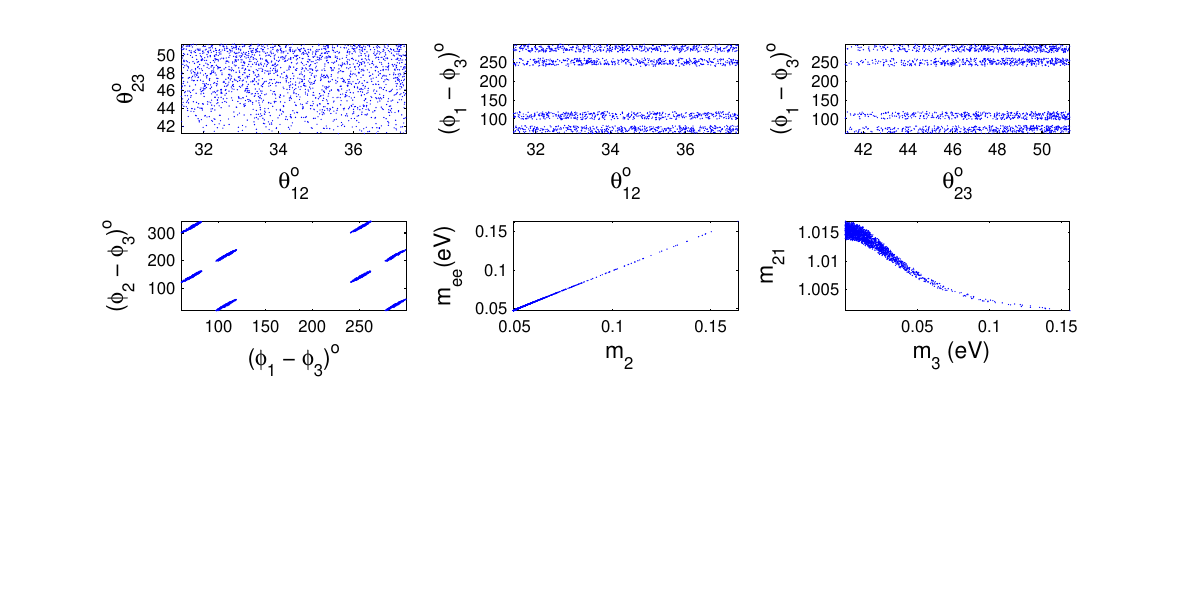}
\caption{The correlation plots for traceless texture in case II (vanishing CP phases), in the inverted ordering hierarchy. }
\label{caseII-inv}
\end{figure}

Fig. (\ref{caseII-inv}) shows the correlations corresponding to inverted hierarchy type.

We note that the mixing angles $\t$
 and the unphysical phases $\phi$ cover all their allowed ranges without gaps. However, the differences ($\phi_1-\phi_3$) and ($\phi_2-\phi_3$) do have disallowed gaps.

We reconstruct the neutrino mass matrix for a representative point, which is taken such that the mixing angles are  near their best fit values from Table \ref{TableLisi:as}. For inverted ordering, the representative point is taken as follows.
\begin{equation}
\begin{aligned}
(\theta_{12},\theta_{23},\theta_{13})=&(34.37^{\circ},49.10^{\circ},8.52^{\circ}),\\
(\phi_1,\phi_2,\phi_3)=&(59.06^{\circ},351.81^{\circ},130.23^{\circ}),\\
(m_{1},m_{2},m_{3})=&(0.0515\textrm{ eV},0.0523\textrm{ eV},0.0151\textrm{ eV}),\\
(m_{e},m_{ee}, \Sigma)=&(0.0512\textrm{ eV},0.0510\textrm{ eV}, 0.1189 \textrm{ eV}),
\end{aligned}
\end{equation}
 and the corresponding neutrino mass matrix (in eV) is
\begin{equation} \label{complexMatrixCPconservedInverted}
M_{\nu}=\left( \begin {array}{ccc} -0.0240+0.0449i &  -0.0024 - 0.0030i &  0.0037 + 0.0006i\\ \noalign{\medskip}-0.0024 - 0.0030i &   0.0301-0.0088i &  0.0095-0.0151i
\\ \noalign{\medskip} 0.0037 + 0.0006i &   0.0095-0.0151i &  -0.0061-0.0361i\end {array} \right).
\end{equation}
Note here that even though  $M_\n$ is complex but no CP violation. Looking at Eqs. (\ref{defOfU}), we see that the equivalent matrix ($P^*_\phi.M_\n.P^*_\phi$), with both unphysical and CP phases equal to zero and hence is real with no phases, does belong to the texture even though it is not traceless, and can not be found if we limited the scanning to real matrices. 

\subsubsection{Normal Hierarchy}
\begin{figure}[hbtp]
\hspace*{-4cm}
\includegraphics[width=24cm, height=16cm]{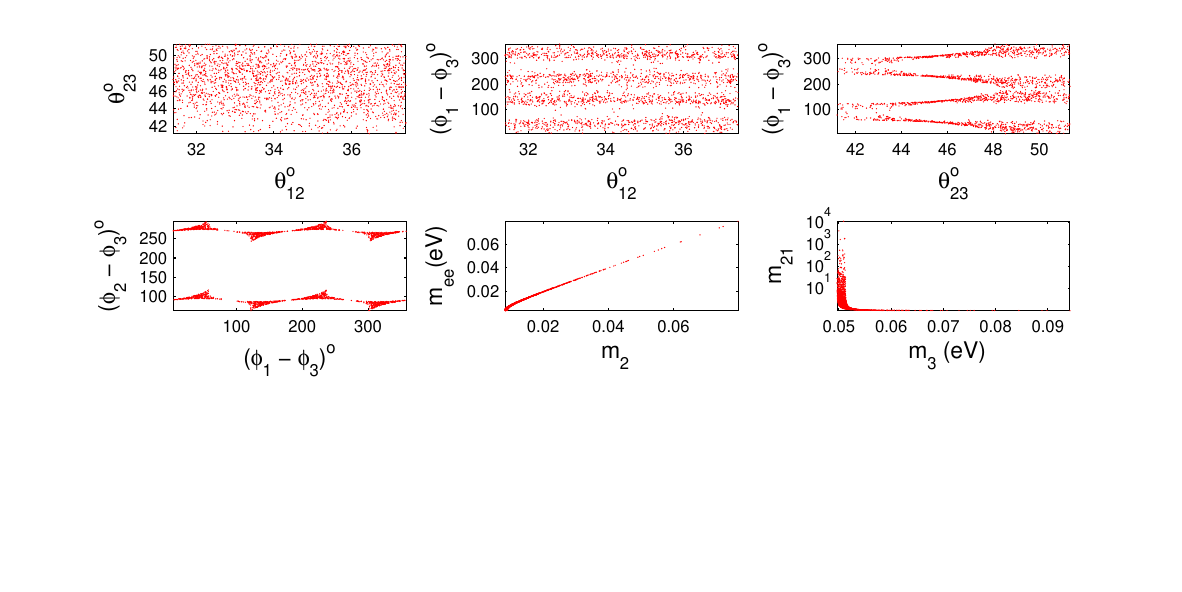}
\caption{The correlation plots for traceless texture in case II (vanishing CP phases), in the normal ordering hierarchy. }
\label{caseII-nor}
\end{figure}

Fig. (\ref{caseII-nor}) shows the correlations corresponding to normal hierarchy type.

We note again that the mixing angles $\t$
 and the unphysical phases $\phi$ cover all their allowed ranges without gaps, and that the differences ($\phi_1-\phi_3$) and ($\phi_2-\phi_3$) do have disallowed gaps.

We reconstruct the neutrino mass matrix for a representative point, which is taken such that the mixing angles are near  from their best fit values from Table \ref{TableLisi:as}. For normal ordering, the representative point is taken as follows.
\begin{equation}
\begin{aligned}
(\theta_{12},\theta_{23},\theta_{13})=&(34.41^{\circ},49.16^{\circ},8.51^{\circ}),\\
(\phi_1,\phi_2,\phi_3)=&(248.68^{\circ},308.20^{\circ},24.07^{\circ}),\\
(m_{1},m_{2},m_{3})=&(0.0164\textrm{ eV},0.0187\textrm{ eV},0.0538\textrm{ eV}),\\
(m_{e},m_{ee}, \Sigma)=&(0.0187\textrm{ eV},0.0179\textrm{ eV}, 0.0889 \textrm{ eV}),
\end{aligned}
\end{equation}
 and the corresponding neutrino mass matrix (in eV) is
\begin{equation} \label{complexMatrixCPconservedNormal}
M_{\nu}=\left( \begin {array}{ccc} -0.0132+0.0121i &  -0.0045 - 0.0014i &  0.0001 - 0.0027i\\ \noalign{\medskip} -0.0045 - 0.0014i & -0.0089-0.0368i & 0.0154-0.0081 i
\\ \noalign{\medskip} 0.0001 - 0.0027i &   0.0154-0.0081 i &  0.0221+0.0247i\end {array} \right).
\end{equation}
Again, here $M_\n$ is complex but with no CP violation. The equivalent real matrix ($P^*_\phi.M_\n.P^*_\phi$) belongs to the texture with no vanishing trace. 

\subsection{General case}
In either hierarchy type, the mixing and phase (CP and unphysical) span their full ranges, with no corresponding gaps. Numerical computations reveal a weak correlations ($\phi_1-\phi_3, \phi_2-\phi_3$), and that $m_{ee}$ is an increasing function of $m_2$. Like in (case II), the inclusion of the unphysical phases dilute many correlations compared to (case I), and so again we present just the six correlations described in (case II).

\subsubsection{Inverted Hierarchy}
\begin{figure}[hbtp]
\hspace*{-4cm}
\includegraphics[width=24cm, height=16cm]{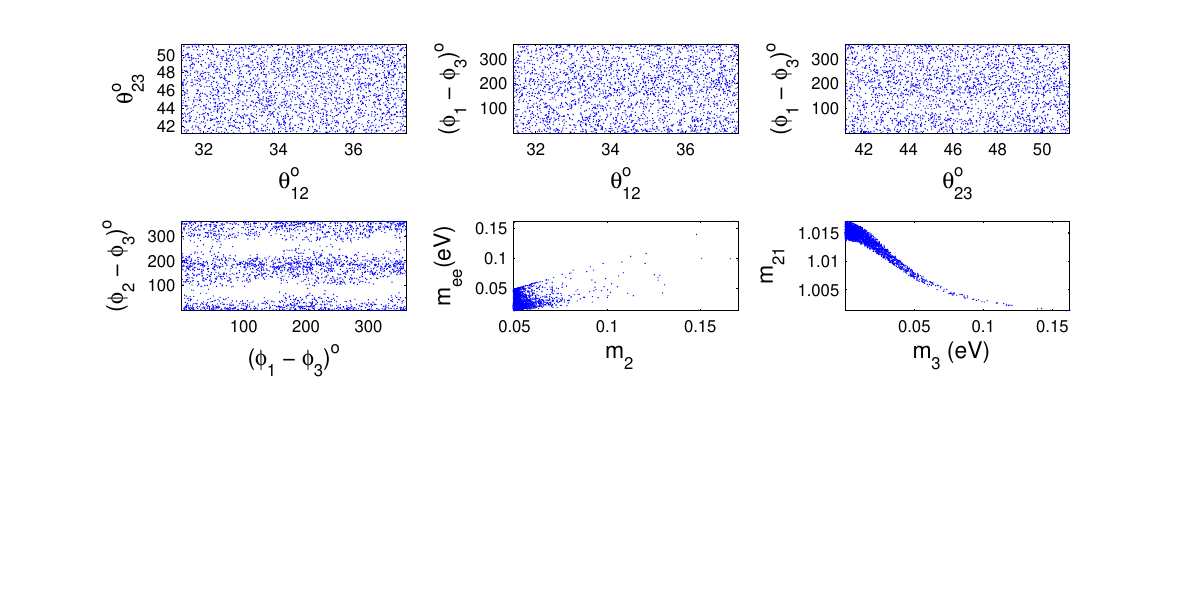}
\caption{The correlation plots for traceless texture in case III (general), in the inverted ordering hierarchy. }
\label{caseIII-inv}
\end{figure}

Fig. (\ref{caseIII-inv}) shows the correlations corresponding to inverted hierarchy type.

Although $m_3$ can reach tiny values of order $m_3 \succeq 2.73 \times 10^{-6}$ eV, but numerically we checked that it can not reach zero exactly, and thus no singular patterns exist.

We reconstruct the neutrino mass matrix for a representative point, which is taken such that the mixing angles and Dirac phase are as close as possible to their best fit values from Table \ref{TableLisi:as}. For inverted ordering, the representative point is taken as follows.
\begin{equation}
\begin{aligned}
(\theta_{12},\theta_{23},\theta_{13})=&(34.82^{\circ},49.36^{\circ},8.51^{\circ}),\\
(\delta,\rho,\sigma)=&(284.60^{\circ},52.50^{\circ},164.81^{\circ}),\\
(\phi_1,\phi_2,\phi_3)=&(39.25^{\circ},334.27^{\circ},326.31^{\circ}),\\
(m_{1},m_{2},m_{3})=&(0.0900\textrm{ eV},0.0904\textrm{ eV},0.0750\textrm{ eV}),\\
(m_{e},m_{ee}, \Sigma)=&(0.0898\textrm{ eV},0.0442\textrm{ eV}, 0.2555 \textrm{ eV}),
\end{aligned}
\end{equation}
 and the corresponding neutrino mass matrix (in eV) is
\begin{equation}
M_{\nu}=\left( \begin {array}{ccc} -0.0397+0.0195i &  0.0488 + 0.0268i &  -0.0451 - 0.0314i\\ \noalign{\medskip}0.0488 + 0.0268i &   0.0147 - 0.0083i &  0.0196-0.0544i
\\ \noalign{\medskip} -0.0451 - 0.0314i &  0.0196-0.0544i &  0.0249-0.0112i\end {array} \right).
\end{equation}

\subsubsection{Normal Hierarchy}
\begin{figure}[hbtp]
\hspace*{-4cm}
\includegraphics[width=24cm, height=16cm]{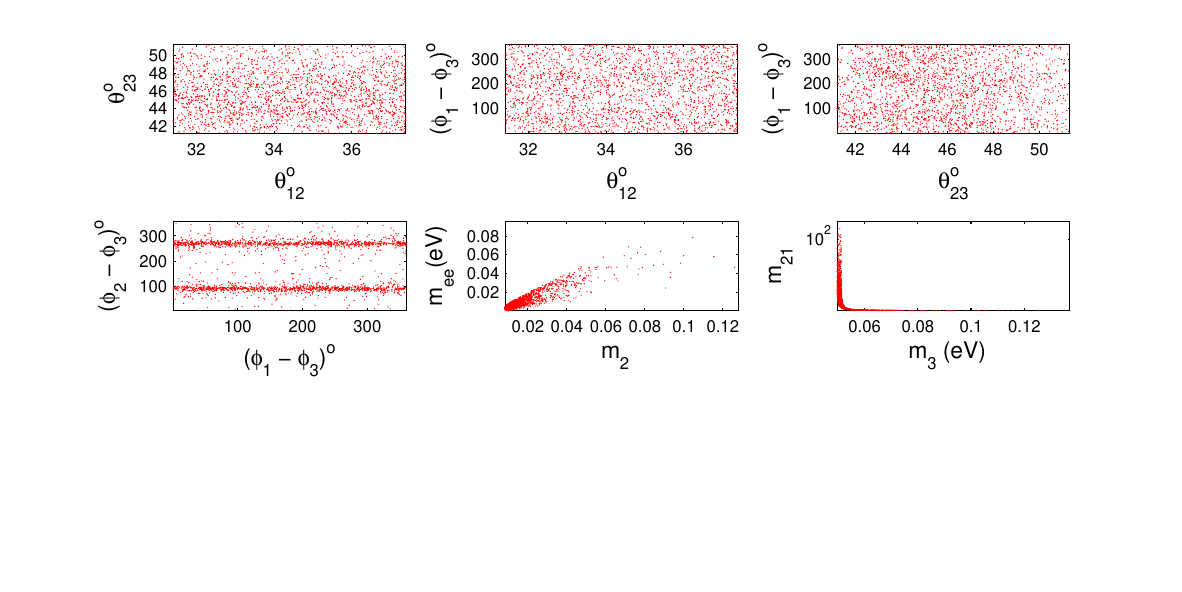}
\caption{The correlation plots for traceless texture in case III (general), in the normal ordering hierarchy. }
\label{caseIII-nor}
\end{figure}

Fig. (\ref{caseIII-nor}) shows the correlations corresponding to inverted hierarchy type.

Although $m_1$ can reach tiny values of order $m_1 \succeq 2.56 \times 10^{-5}$ eV, but numerically we checked that it can not reach zero exactly, and thus no singular patterns exist.

We reconstruct the neutrino mass matrix for a representative point, which is taken such that the mixing angles and Dirac phase are as close as possible to their best fit values from Table \ref{TableLisi:as}. For normal ordering, the representative point is taken as follows.
\begin{equation}
\begin{aligned}
(\theta_{12},\theta_{23},\theta_{13})=&(34.33^{\circ},49.42^{\circ},8.55^{\circ}),\\
(\delta,\rho,\sigma)=&(284.74^{\circ},170.22^{\circ},95.82^{\circ}),\\
(\phi_1,\phi_2,\phi_3)=&(32.16^{\circ},119.59^{\circ},45.10^{\circ}),\\
(m_{1},m_{2},m_{3})=&(0.0314\textrm{ eV},0.0327\textrm{ eV},0.0592\textrm{ eV}),\\
(m_{e},m_{ee}, \Sigma)=&(0.0327\textrm{ eV},0.0143\textrm{ eV}, 0.1233 \textrm{ eV}),
\end{aligned}
\end{equation}
 and the corresponding neutrino mass matrix (in eV) is
\begin{equation}
M_{\nu}=\left( \begin {array}{ccc} 0.0130+0.006i &  -0.0019 + 0.0199i &  -0.0213 + 0.0029i\\ \noalign{\medskip}  -0.0019 + 0.0199i  & -0.0088-0.0341i & -0.0240+0.0146 i
\\ \noalign{\medskip}-0.0213 + 0.0029i &  -0.0240+0.0146 i &  -0.0041+0.0280i\end {array} \right).
\end{equation}

\subsection{Number of free parameters}
As said, a consistent rephasing-invariant definition should be defined on the set of equivalence classes $\mathcal{M}/\sim$, which does not have a vector space structure, so the number of free parameters (degrees of freedom)\footnote{We mean the ``minimal" number of free parameters, where -by assumption- each parameter covers an interval of real values.} accounting for a texture is used rather than a ``dimension" of the texture space. Put it another way, one can use the freedom of 3 arbitrary phases, so that $\mathcal{M}/\sim$ can be accounted for by 9 real parameters. 

In cases I (II), the phases freedom was used in picking up a slice of vanishing unphysical (CP) phases, parameterized thus by 9 parameters. Now, imposing the traceless condition amounts to two real constraints, so both cases can be accounted for by 7 free parameters, in line with our use of 7 parameters in the corresponding scanning. 

The case III, where we do not restrict the study to a slice but rather impose the traceless condition then we count the degrees of freedom, is more subtle. While the vector space of traceless symmetric complex matrices is 10-dimensional, in line with our use of 10 parameters when scanning, many equivalent matrices count for one texture realization, so the question arises as to what is the  number of free parameters accounting for the `traceless' equivalence classes $({\mathcal{M}/\sim})^{\mbox{\tiny traceless}}$ (which is $\leq 9$ by using the phases freedom and $\geq 7$ from cases I \& II) corresponding to case III\footnote{A `traceless' equivalence class is an equivalence class with one traceless representative matrix.} ?  

One might be tempted to say that  exhausting the freedom of phases after imposing the traceless condition  would give again  ($10-3=7$) free parameters. This is incorrect due to the fact that the traceless condition is not independent from the unphysical phases, and one can not treat them separately. Actually, once we restrict ourselves to the 10-dim submanifold of traceless symmetric matrices, we do not have full freedom of phasing, as multiplying from left and right by a phase matrix may lead to exiting the traceless submanifold. 

Actually, using the rephasing freedom in order to absorb the diagonal elements phases, one can pick up a particular 9-parameter slice $S_1$ of complex symmetric matrices characterized by non-zero real diagonal elements with signs $++-$, where forcing the traceless condition corresponds now to one degree of freedom and hence would amount to reducing the number of parameters to $8$ accounting for the set of inequivalent matrices of the form
\bea \label{tracelessS1}
\left\{ 
\left[ \left ( \begin{array}{ccc} a  &  b   &  c \\ b & d 
& e \\ c  & e
 & -a-d \end{array} \right ) \right]
 : a,d \in \mathbf{R}^{+*}, b,c,e \in \mathbf{C}
 \right\}
 \eea   
Hence at least we need 8 real parameters to cover the traceless equivalence classes $({\mathcal{M}/\sim})^{\mbox{\tiny traceless}}$ meeting the traceless condition according to ``generalized" definition. 

We see explicitly here that the number of free parameters in the ``specific" definition (Eq. \ref{specific}) depends on the slice $S$ used in the definition. Actually, contrasting the slice $S_1$ with those in cases I \& II, we see that we started by picking up a 9-parameter slice, then we imposed the traceless condition, but while it happens that for the slice under consideration in cases I \& II the trace was complex corresponding to 2 real degrees of freedom, the trace in ($S_1$) is real and the traceless condition corresponded only to one degree of freedom.  

Note that there are matrices meeting the ``generalized" traceless condition which can not be recovered from $S_1$, such as any matrix $G$ whose diagonal elements are $\left( 1,1,1\right)$. Clearly such a matrix does not satisfy the ``mathematical" definition. Moreover, its equivalent representative in the slice $S_1$, obtained by multiplication from right and left by the matrix $\mbox{diag} \left(1,1,i\right)$, would be a matrix whose diagonal elements are $\left( 1,1,-1\right)$, which again does not satisfy the traceless condition. However $G$ does meet the ``generalized" traceless condition, and actually by multiplying it from right and left by the phase matrix  $\mbox{diag} \left(1,\sqrt{j},j\right):j=e^{\frac{2i\pi}{3}}$, we get the equivalent matrix $G^\prime$ of diagonal elements  $\left( 1,j,j^2\right)$ meeting the traceless condition. One cannot equate the two ``positive" parameters ($(a,d)\in R^{{+*}^2}$), in Eq. (\ref{tracelessS1}), with one ``real" parameter, so the parameter space for the ``traceless generalized definition" is at most 9, while  8 parameters are not enough to account for it.

In order to find exactly the number of free parameters, it is useful to find a ``good" definition,  i.e. manifestly rephasing-invariant, of the ``generalized" definition (Eq. \ref{good}). In appendix 1, we find such a definition given by the following inequality involving moduli of $M_\n$ elements unaffected by unphysical phases:
\bea
\label{generalized traceless1}
|M_{\n 11}|^4 + |M_{\n 22}|^4 + |M_{\n 33}|^4 \leq 2\left( |M_{\n 11}|^2 |M_{\n 22}|^2 + |M_{\n 11}|^2 |M_{\n 33}|^2 + |M_{\n 22}|^2 |M_{\n 33}|^2 \right)
\eea         
Since we have an inequality rather than an equality, then the number of free parameters remains $9$ but the range of one parameter is further constrained via the inequality. In Appendix 2, we give a pictorial representation of the various definitions illustrating as well the question of the number of free parameters.


\section{Theoretical realization of case III}
We present now one realization of the traceless texture based on $A_5$ non-abelian flavor symmetry within seesaw type II scenario.
Although the matter content is extended beyond SM, to include new scalars, however we shall not discuss the question of the scalar potential and finding its general form under the imposed symmetry giving the required vacuum expectation values (VEVs). Actually, with new scalars, rich phenomenology at colliders may arise, and requesting only one SM-like Higgs at low scale is not a trivial task, and requires generally fine tuning.

\subsection{$A_5$ group theory}
The alternating group $A_5$ consists of even permutation between five elements. It is finite containing ($n_G=60$) elements which can be classified into five conjugation classes\footnote{The conjugation relation defined by ($x \equiv y \Leftrightarrow \exists z \in G: x= z.y.z^{-1}$) is an equivalence relation partitioning the group into conjugation classes.}. Thus there are five independent unitary irreducible representations (irreps), which we call ($\bf 1, 3, 3', 4, 5$). We state now the multiplication rules for these irreps (Eq. \ref{a5irreps}), and put also the SALC (symmetry adapted linear combinations) which would serve us in constructing the model when multiplying two triplets (Eq. \ref{3x3}), with the singlet combination when multiplying two quantuplets (Eq. \ref{5x5}).
\bea
\label{a5irreps}
{\bf 3} \otimes {\bf 3} = {\bf 1} \oplus {\bf 3} \oplus {\bf 5}
&,&
 {\bf 3'} \otimes {\bf 3'} = {\bf 1} \oplus {\bf 3'} \oplus {\bf 5},
 \\
 {\bf 3} \otimes {\bf 3'} = {\bf 4} \oplus {\bf 5}
&,&
{\bf 4} \otimes {\bf 5} = {\bf 3} \oplus {\bf 3'}  \oplus  {\bf 4} \oplus  {\bf 5} \oplus  {\bf 5},
\nn\\
{\bf 3} \otimes {\bf 4} = {\bf 3'}\oplus {\bf 4} \oplus {\bf 5}
&,&
{\bf 3'} \otimes {\bf 4} = {\bf 3}\oplus {\bf 4} \oplus {\bf 5}
\nn\\
{\bf 3} \otimes {\bf 5} =  {\bf 3}\oplus {\bf 3'}\oplus {\bf 4} \oplus {\bf 5}
&,&
{\bf 3'} \otimes {\bf 5} =  {\bf 3}\oplus {\bf 3'}\oplus {\bf 4} \oplus {\bf 5}
\nn\\
 {\bf 4} \otimes {\bf 4} = {\bf 1} \oplus {\bf 3} \oplus {\bf 3'} \oplus {\bf 4}\oplus {\bf 5}
 &,&
 {\bf 5} \otimes {\bf 5} = {\bf 1} \oplus {\bf 3} \oplus {\bf 3'}   \oplus {\bf 4} \oplus {\bf 4}\oplus {\bf 5} \oplus {\bf 5},
 \nn \\
\left(\begin{array}{l}
a_{1} \\
a_{2} \\
a_{3}
\end{array}\right)_{\bf 3} \otimes\left(\begin{array}{l}
b_{1} \\
b_{2} \\
b_{3}
\end{array}\right)_{\bf 3}&=& \left(a_{1} b_{1}+a_{2} b_{2}+a_{3} b_{3}\right)_{\bf 1} \oplus \left(\begin{array}{c}
a_{3} b_{2}-a_{2} b_{3} \\
a_{1} b_{3}-a_{3} b_{1} \\
a_{2} b_{1}-a_{1} b_{2}
\end{array}\right)_{\bf 3} \nn \\
 && \oplus\left(\begin{array}{c}
a_{2} b_{2}-a_{1} b_{1} \\
a_{2} b_{1}+a_{1} b_{2} \\
a_{3} b_{2}+a_{2} b_{3} \\
a_{1} b_{3}+a_{3} b_{1} \\
-\frac{1}{\sqrt{3}} \left( a_{1} b_{1}+a_{2} b_{2}-2 a_{3} b_{3}\right)
\end{array}\right)_{\bf 5}, \label{3x3} \\
\left(a_{1}, a_{2}, a_{3}, a_4, a_5 \right)^T_{\bf 5}
 \otimes\left(b_{1},b_{2}, b_{3}, b_{4}, b_{5} \right)^T_{\bf 5} &\supset& \left( a_1 b_1 + a_2 b_2 + a_3 b_3 + a_4 b_4 + a_5 b_5 \right)_{\bf 1} \label{5x5}.
\eea
For more details on $A_5$ group, one can consult \cite{Ishimori_2010}.

\subsubsection{Type-II Seesaw Matter Content:}
\label{subsubsection-a5-matter}
For the type-II seesaw scenario leading to a traceless texture of the neutrino mass matrix, we take the matter content shown in Table (\ref{matter-content-A5})

 \begin{table}[h]
\caption{matter content and symmetry transformations, leading to traceless texture. $\Delta_j= \left( \begin{array} {cc} \Delta^+_j &  \Delta^{++}_ j \\  \Delta^{0}_ j & -\Delta^+_j  \end{array}\right)$, $j=1,\ldots,5$ with $i=1,2,3$ is a family index.}
\centering
\begin{tabular}{ccccccc}
\hline
\hline
Fields & $D_{L_i}$ & $\Delta_j$ &  $\ell_{R_i}$ & $\phi_I$ & $\phi_{{III}_i}$ & $\phi_{V_j}$    \\
\hline
\hline
$SU(2)_L$ & 2 & 3 & 1 & 2 & 2& 2  \\
\hline
$A_5$ & ${\bf 3}$ & ${\bf 5}$ & ${\bf 3}$   & ${\bf 1}$ & ${\bf 3}$ & ${\bf 5}$
 \\
\hline
\end{tabular}
\label{matter-content-A5}
\end{table}
The Lorentz-, gauge- and $A_5$-invariant terms relevant for the neutrino mass matrix are
\bea
\label{Lagrangian-neutrino-a5}
{\cal L} \ni \left( D_L^T D_L \Delta \right)_{\bf 1} &=& Y  \left[
\left( D^T_{L\mu} C^{-1} i \tau_2  D_{L\m} - D^T_{Le} C^{-1} i \tau_2  D_{Le} \right) \Delta_1 \nn \right. \\ &&
+  \left( D^T_{L\mu} C^{-1} i \tau_2  D_{Le} + D^T_{Le} C^{-1} i \tau_2  D_{L\m} \right) \Delta_2 \nn \\ &&
+\left( D^T_{L\tau} C^{-1} i \tau_2  D_{L\m} + D^T_{L\m} C^{-1} i \tau_2  D_{L\tau} \right) \Delta_3 \nn \\ &&
+\left( D^T_{Le} C^{-1} i \tau_2  D_{L\tau} + D^T_{L\tau} C^{-1} i \tau_2  D_{Le} \right) \Delta_4 \nn \\ &&
-\frac{1}{\sqrt{3}}\left( D^T_{Le} C^{-1} i \tau_2  D_{Le} + D^T_{L\m} C^{-1} i \tau_2  D_{L\m} - 2 D^T_{L\tau} C^{-1} i \tau_2  D_{L\tau} \right) \Delta_5 \left.
\right]
\eea
This $Y$-term picks up the  quintuplet combination from the product of the two triplets ($D^T_{L_i}$ and $D_{L_i}$) (Eq. \ref{3x3}), before multiplying it with the Higgs flavor quintuplet $\Delta_j$ in order to get a singlet according to Eq. (\ref{5x5}). We get, upon acquiring small VEVs for the neutral components of $\Delta_j^o, j=1,\ldots,5$, denoted by $v_j$ and set by the Lagrangian free parameters\footnote{We do not discuss how the minimization of the scalar potential leads to VEVs of small order such that when multiplied by the Yukawa perturbative coupling give the right order of magnitude for the neutrino mass. However, we note that these VEVs should be quite small compared to the electroweak scale due to the heavy triplet mass terms.}, the following neutrino mass matrix
\bea
\label{realization_neutrino}
M_{\n} &=& Y \left( \begin{array}{ccc}-\frac{1}{\sqrt{3}} v_5 - v_1&v_2&v_4\\v_2&-\frac{1}{\sqrt{3}} v_5 + v_1&v_3\\v_4&v_3&\frac{2}{\sqrt{3}}v_5\end{array}\right)
\eea
We see that this texture meets the traceless condition $\left[\mbox{Tr}(M_\n)=0\right]$.

\subsubsection{Charged lepton sector:}
\label{subsubsection-a5-charged}
We  can build a generic mass matrix and impose suitable hierarchy conditions in order to diagonlize $M_\ell$ by rotating infinitesimally the left-handed charged lepton fields. This means that, up to approximations of the order of the charged lepton mass-ratios hierarchies, we are in the `flavor' basis, and the previous phenomenological study is valid. These corrections due to rotating the fields are not larger than other, hitherto discarded, corrections coming, say, from radiative renormalization group running from the seesaw high scale to the observed data low scale.

The Lorentz-, gauge- and $A_5$-invariant terms relevant for the charged lepton mass are
\bea
\label{Lagrangian-chargedlepton-a5}
{\cal L} \ni \left(\Sigma_{a=I, III, V}\overline{D_L} \ell_R \;\phi_a\right)_{\bf 1} &=& Y_I  \left( \overline{D}_{Le} e_R + \overline{D}_{L\m} \m_R + \overline{D}_{L\tau} \tau_R  \right) \phi_I  \nn \\  && +
 Y_{III}  \left[\right. \left( \overline{D}_{L\tau} \m_R - \overline{D}_{L\m} \tau_R\right) \phi_{III_1}
 +\left( \overline{D}_{Le} \tau_R - \overline{D}_{L\tau} e_R\right) \phi_{III_2} \nn \\ &&
+\left( \overline{D}_{L\m} e_R - \overline{D}_{Le} \m_R\right) \phi_{III_3} \left. \right]\nn \\ &&
 +Y_{V}  \left[ \left( \overline{D}_{L\m} \m_R - \overline{D}_{Le} e_R\right) \phi_{V_1} \right.
+\left( \overline{D}_{L\m} e_R + \overline{D}_{Le} \m_R\right) \phi_{V_2} \nn \\ &&
+\left( \overline{D}_{L\tau} \m_R + \overline{D}_{L\m} \tau_R\right) \phi_{V_3}
+\left( \overline{D}_{Le} \tau_R + \overline{D}_{L\tau} e_R\right) \phi_{V_4} \nn \\ &&
-\frac{1}{\sqrt{3}} \left( \overline{D}_{Le} e_R + \overline{D}_{L\m} \m_R - 2 \overline{D}_{L\tau} \tau_R \right) \phi_{V_5} \left. \right]
\eea
which leads, when $\phi_a, a\in\{I,III,V\}$ acquires a VEV, to a  charged lepton mass:
\bea
\label{charged-Lepton-mass-matrix-A5}
M_{\ell} &=& Y_I \langle \phi_{I}\rangle_0 \left( \begin{array}{ccc}1&0&0\\0&1&0\\0&0&1\end{array}\right)
+Y_{III} \left( \begin{array}{ccc}0&-\langle \phi_{III_3}\rangle_0&\langle \phi_{III_2}\rangle_0\\\langle \phi_{III_3}\rangle_0&0&-\langle \phi_{III_1}\rangle_0\\-\langle \phi_{III_2}\rangle_0&\langle \phi_{III_1}\rangle_0&0\end{array}\right)  \nn \\ &&+
Y_V \left( \begin{array}{ccc}-\frac{1}{\sqrt{3}}\langle \phi_{V_5}\rangle_0-\langle \phi_{V_1}\rangle_0&\langle \phi_{V_2}\rangle_0&\langle \phi_{V_4}\rangle_0\\\langle \phi_{V_2}\rangle_0&-\frac{1}{\sqrt{3}}\langle \phi_{V_5}\rangle_0+\langle \phi_{V_1}\rangle_0&\langle \phi_{V_3}\rangle_0 \\\langle \phi_{V_4}\rangle_0&\langle \phi_{V_3}\rangle_0& \frac{2}{\sqrt{3}}\langle \phi_{V_5}\rangle_0 \end{array}\right).
\eea
Two common ways to get a generic $M_\ell$.
\begin{itemize}
\item We assume a VEV hierarchy such that the $(\langle \phi_{I}\rangle_0, \langle \phi_{V_1}\rangle_0, \langle \phi_{V_5}\rangle_0)$ are dominant , and we get approximately a diagonal matrix
    \bea
    \label{chargedMass-dominance}
    M_{\ell} &\approx& \mbox{diag} \left(M_{\ell 11}, M_{\ell 22}, M_{\ell 33} \right): \\
    M_{\ell 11} &=& Y_I \langle \phi_{I}\rangle_0 -\frac{1}{\sqrt{3}} Y_V \langle \phi_{V_5}\rangle_0- Y_V \langle \phi_{V_1}\rangle_0, \nn \\
    M_{\ell 22} &=& Y_I \langle \phi_{I}\rangle_0 -\frac{1}{\sqrt{3}}Y_V \langle \phi_{V_5}\rangle_0+ Y_V \langle \phi_{V_1}\rangle_0, \nn \\
    M_{\ell 33} &=& Y_I \langle \phi_{I}\rangle_0  + \frac{2}{\sqrt{3}} Y_V \langle \phi_{V_5}\rangle_0. \nn
    \eea
    With the free two Yukawa couplings and three VEVs, one can accommodate the observed charged lepton masses. Thus, as long as the neglected VEVs are sufficiently small, the rotation into the `flavor' basis is done via infintesimal rotations.

\item Looking at Eq. (\ref{charged-Lepton-mass-matrix-A5}), we see that we have 9 free VEVs and 3 free perturbative coupling constants, appearing in 9 linear independent combinations, {\it a priori} enough to construct the generic $3 \times 3$ complex matrix. Thus,  $M_\ell$ can be casted in the form
   \bea
\label{M_elltype2}
M_{\ell } =  \left( \begin {array}{c}
{\bf a}^T\\{\bf b}^T\\{\bf c}^T
\end {array}
\right) &\Rightarrow&
M_{\ell } M_{\ell}^\dagger = \left(\begin {array}{ccc}
{\bf a.a} &{\bf a.b}&{\bf a.c} \\
{\bf b.a} &{\bf b.b}&{\bf b.c}\\
{\bf c.a} &{\bf c.b}&{\bf c.c}
\end {array} \right)
\eea
where ${\bf a}, {\bf b}$ and ${\bf c}$ are three linearly independent vectors, so taking only the following natural assumption on the norms of the vectors
\bea \parallel {\bf a} \parallel /\parallel {\bf c} \parallel = m_e/m_\tau \sim 3 \times 10^{-4} &,&  \parallel {\bf b} \parallel /\parallel {\bf c} \parallel = m_\mu/m_\tau \sim 6 \times 10^{-2}\eea
one can diagonalize $M_{\ell } M_{\ell}^\dagger$ by an infinitesimal rotation as was done in \cite{0texture}, which proves that we are to a good approximation in the flavor basis.

\end{itemize}

\section{Summary and Conclusion}
In this study, we have revisited the traceless texture for neutrino mass matrix, and have carried out a systematic study highlighting the role played by all phases, with a focus on the correlations, resulting from imposing the traceless condition, between the physical experimentally observable parameters and the unphysical phases which can be absorbed by the charged lepton fields. Although the unphysical phases are devoid of observable physics in SM augmented by neutrino masses, however one needs to take them into consideration in order to give a consistent texture definition, in a similar way that one should keep them when developing the renormalization group equations for the neutrino masses and mixing angles at different energy scales. 

We studied three cases which correspond actually to three different textures, and so three resulting phenomenologies, although apparently the difference between the first and third cases lies only in including or droping the unphysical phases, which would have suggested similar phenomenolgies, were it not for the fact that introducing the unphysical phases in a consistent way does change the definition of the texture, whence leading to different phenomenology.

Case I, matching all previous studies of traceless texture and corresponding to switching off the unphysical phases, leads to acceptable traceless texture accommodating both types of hierarchy. This case, as well as the previous studies assuming vanishing unphysical phases, should be looked at as being corresponding to a different texture from when switching on these phases, otherwise one could not a priori limit the study to the vanishing unphysical phases slice. 

Moreover, this case corresponds to a definition which is dependent on the PMNS parametrization one chooses. Taking this last remark into consideration, switching between different parametrizations, our results, apart from slight changes due to updating the experimental data, are in line, when taking the PDG parameterization and using the same scanning strategy described in Sec.~4, with the past study in \cite{Rodejohann2004}, which assumed vanishing unphysical phases in the PDG parametrization. We opted not to list the corresponding numerical results and plots in order to avoid unnecessary confusion. As a consistency check, we imposed the traceless condition in the PDG parametrization within the submanifold ($\phi'_1=-\beta, \phi'_2=\phi'_3= - \beta - \delta$) so that to produce our case I plots, as it should be according to the transformation laws of Eq.~(\ref{param}), and indeed we got the same  correlation plots and numerical results. 

Restoring the unphysical phases while turning off the CP phases (case II) allows also for both hierarchy types, and the definition of the texture in this case is the same in both Aopted and PDG parametrizations, as vanishing CP phases in the former ($\d, \r, \s$)   corresponds well to them ($\d, \a, \b$) vanishing in the latter. Case III assumed both unphysical and CP phases existent, and gave correlations with the most accepted points. The definition here , according to the generalized one,  is both ``rephasing''invariant and parametrization independent. 

Actually,  including the unphysical phases, we find that many correlations get ``diluted" (voire ``disappeared''), since, with these unphysical phases not necessarily zero, one can find many more points in the parameter space meeting the experimental constraints, for the observable parameters, and the mathematical traceless condition, decreasing thus the predictive power of the texture compared to previous studies. One should comment that upon introducing unphysical phases then the different parameterizations, related to each other by certain relations \cite{chamoun2023}, give the same correlations, when adjusted to correspond to the same definitions, and same phenomenology, as was explicitly checked by us.

Finally, we presented a theoretical realization of these textures via a type-II seesaw scenario and assuming a non-abelian group $A_5$.    However, we have not discussed the question of the scalar potential
and finding its general form under the imposed symmetry. Nor did we deal with the radiative corrections effect on the phenomenology and whether or not it can spoil the form of the texture while running from the seesaw “ultraviolet” scale, where the traceless condition is imposed, to the low scale where phenomenology was analyzed.

{
We end the paper by recapitulating the main points of the strategy, related to unphysical phases, that one should follow when studying textures of the neutrino mass matrix. 
\begin{enumerate}
\item Since the unphysical phases are non physical, then any physically consistent definition of any texture should be rephasing-invariant ($\phi^{\mbox{\tiny unphysical}}-$invariance), in that any two equivalent matrices, obtained one from the other by multiplying from left and right with a phase matrix, should either both belong to the texture or neither does.  
\item For many textures (like zero-textures, vanishing minor textures, zero-determinant textures, and many others) rephasing-invariance is automatic, but for others, like the traceless texture, albeit relatively simple, the ``mathematical definition" is not rephasing-invariant, and the correct way to define the texture is to use the ``generalized definition" (e.g. case III for the traceless texture), in that a matrix belongs to the texture if and only if one equivalent matrix does satisfy the mathematical definition. 
   The ``mathematical" and ``generalized" definitions are parametrization independent, but only the latter is rephasing-invariant in general.
   \item  Finding out matrices meeting either of the ``mathematical" or ``generalized" definitions amounts to scanning over $M_\n$-parameters including the unphysical phases, and although the two definitions are different but they lead to the same phenomenology. It is more illuminating to find a ``good mathematical" definition of the ``generalized definition" dropping manifestly any dependence on the unphysical phases, but this is not guaranteed in general. Furthermore, these ``good generalized" definitions often involve non-analytic functions, such as modulus or argument, which would make the symmetry realization exercise rather difficult.   
\item Some past studies were done by imposing the ``mathematical constraint" on a particular slice, namely the vanishing unphysical phases slice. In order to make these studies physically consistent, we introduced the ``specific definition" which thus becomes rephasing-invariant. According to this definition, a matrix belongs to the texture if and only if its equivalent matrix with vanishing unphysical phases does satisfy the mathematical condition.   
\item However the ``Specific definition" depends on the parametrization, because it singles out, in the parameter space, the slice of vanishing unphysical phases, which changes from parametrization to another,  as the slice where to impose the mathematical constraint (e.g. case I for the traceless texture, where the resulting phenomenologies are different in general when one adopts different parametrizations).  
\item One can choose another slice in the parameter space where to impose the mathematical constraint. A good choice amounts to it being parametrization-independent. In case II, we imposed the mathematical constraint on the slice of vanishing CP-phases. Again within the philosophy of ``specific definition'', one can define that a matrix belongs to the texture if and only if its equivalent matrix with vanishing CP-phases satisfies the mathematical constraint. In addition to be rephasing-invariant, and since the slice of vanishing CP-phases is the same irrespective of the parametrization chosen (it is the slice of CP conservation), so the definition here is also parametrization independent.
\item For Flavor model building, and since it is difficult to check that the resulting $M_\n$ lies in the particular slice on which we imposed the `mathematical' constraint, so it amounts in general to build the texture in the sense of the ``generalized definition".       
\end{enumerate}
}

\section*{{\large \bf Acknowledgements}}
E. I. L acknowledges support from ICTP through the Senior Associate program and the short-term visits. N. C. acknowledges support from the CAS PIFI fellowship and from the Humboldt Foundation. 
The work of E. I. L is partially supportrd by the Science, Technology \& Innovation Funding Authority (STDF) under grant number 50806. 

\appendix
{
\paragraph{Appendix 1: The generalized traceless texture}
The ``mathematical definition" of the traceless texture for $M_\n$ is:
\bea \label{traceless1}\sum_{i=1}^{i=3} M_{\n ii} = 0\eea
The corresponding ``generalized definition" amounts to: 
\bea \label{GT 11}
\exists \phi_1, \phi_2, \phi_3 &:& e^{2i \phi_1}M_{\n 11}+ e^{2i \phi_2}M_{\n 22}+ e^{2i \phi_3}M_{\n 33} =0
\eea  
One might seek a ``good" mathematical definition equivalent to this ``generalized definition" which is manifestly ``rephasing-invariant". 
For this we write (Eq. \ref{GT 11}) as:
\bea 
\left|M_{33}\right|^2=\left|M_{11}+e^{i \alpha} M_{22}\right|^2&:& \alpha=2\left(\phi_2-\phi_1\right) \Rightarrow \nn \\
\left|M_{33}\right|^2-\left|M_{11}\right|^2-\left|M_{22}\right|^2=2 \operatorname{Re}\left(\left|M_{11}\right|\left|M_{22}\right| e^{i \beta}\right)&:& \beta=-\alpha+\arg \left(M_{11}\right)-\arg \left(M_{22}\right)\nn
\eea
and so we get
\bea \label{generalized traceless1}
\left\{\begin{array}{ll}
M_{11}=0 \Rightarrow & \left|M_{33}\right|=\left|M_{22}\right| \\
M_{22}=0 \Rightarrow & \left|M_{33}\right|=\left|M_{11}\right| \\
M_{11} M_{22} \neq 0 \Rightarrow & \frac{\left|M_{33}\right|^2-\left|M_{11}\right|^2-\left|M_{22}\right|^2}{\left|M_{11}\right|\left|M_{22}\right|}=2 \cos \beta \Rightarrow\left|\frac{\left|M_{33}\right|^2}{\left|M_{11}\right|\left|M_{22}\right|}-\frac{\left|M_{11}\right|}{\left|M_{22}\right|}-\frac{\left|M_{22}\right|}{\left|M_{11}\right| }\right| \leq 2
\end{array}\right.
\eea
i.e. we have the following ``good generalized traceless" condition:
\bea
\label{generalized traceless1_1}
|M_{\n 11}|^4 + |M_{\n 22}|^4 + |M_{\n 33}|^4 \leq 2\left( |M_{\n 11}|^2 |M_{\n 22}|^2 + |M_{\n 11}|^2 |M_{\n 33}|^2 + |M_{\n 22}|^2 |M_{\n 33}|^2 \right)
\eea
The constraint of (Eq. \ref{generalized traceless1_1}) expresses the ``generalized traceless texture" in a form manifestly $\phi^{\mbox{\tiny unphysical}}-$invariant. The two constraints of (Eq. \ref{traceless1}) and (Eq. \ref{generalized traceless1_1}) are clearly different, but they give the same phenomenology as regards the observable correlations dropping the unphysical phases.  
}

{
\paragraph{Appendix 2: Pictorial Representation}

Pictorially, one can depict the various definitions by representing $\cal M$, the $12$-dim complex symmetric matrices, by the 2-dim polar punctured plane (Fig. \ref{schema}), with the ``generalized radius $R=\sqrt{X^2+Y^2}$" (angle) coordinate describing the unphysical phases (remaining physical parameters), and where the slice of vanishing unphysical phases is portrayed as the ``generalized unit-circle" of equation ($X^2 + Y^2 =1$). However, each parametrization ($X, Y$) corresponds to a basis vector on the y-axis of length given by $s$, i.e. $X=x, Y=\frac{y}{s}$ and so in polar coordinates the unphysical phases is given by ($R= \sqrt{X^2+Y^2}=\sqrt{x^2 + (\frac{y}{s})^2}$ whereas the observables are given by $\theta = \arctan{\frac{y}{x}}=\arctan{\frac{sY}{X}}$), and let us choose $s=(>)1$ for Adopted (PDG) parametrization. The various vanishing unphysical phases slices are given thus by ellipses with demi-axes ($1,s$) in the $(x,y)$ coordinate system. Then, experimental constraints, as well as the points meeting the `Generalized' (`Specific') definitions,  correspond to an union of radials. Real matrices, having vanishing phases regardless of the parametrization, belong to the intersection of vanishing unphysical phases slices, i.e. points ($A,B$). The CP-conserved slice is parametrization independent so should be union of radials as well, including the real matrices ($A,B$). 

As an example, meant to illustrate the various notions with no claim to proof, let us assume, say, the following (sectors are spanned in anticlockwise direction):
\bea
&\mbox{Experimental constraints are sector DOE}, \nn\\
&\mbox{Traceless matrices ($g=0$) are black curve asymptote of radial OC: radials define sector COA},\nn\\
&\mbox{CP conservation slice is the x-axis (union of radials OA and OB)} \nn
\\ &\mbox{$A$ ($B$) represents real matrices with zero (non-zero) trace. $A' \subseteq A$ with diagonal signs $++-$,} \nn \\
&\mbox{Complex matrices with real non-vanishing diagonal with $++-$ signs ($S_1$) are ``punctured" rectangle,} \nn \\&\mbox{Two holes on rectangle represent matrices with real diagonal elements containing a zero} \nn
\eea
Then, answering the question about number of parameters for $\left(\mathcal{M}/\sim\right)^{\mbox {\tiny traceless}}$ represented by radials of the sector $COA$,  we have
\bea
&\mbox{`Mathematical' definition points: part of `black' curve in the sector COE}, \nn\\
&\mbox{Case I Adopted: radial OH (intersection of traceless \& circle gives \{H,A\}. H passes tests)}, \nn \\
&\mbox{Case I PDG: radial OJ (intersection of traceless \& ellipsis gives \{J,A\}. J passes tests)}, \nn \\
&\!\!\!\!\!\!\!\!\!\!\!\!\!\!\!\!\!\!\!\!\!\!\!\!\!\!\!\!\mbox{Case II: radial OL (intersection of traceless \& CP-conserved gives \{L,A\}. L (`complex': Eqs.(\ref{complexMatrixCPconservedInverted},\ref{complexMatrixCPconservedNormal})) passes tests),} \nn \\
&\mbox{Case III: is sector COE (traceless radials span sector COA but sector COE passes tests)}, \nn \\
&\mbox{Traceless matrices with real non-vanishing diagonal with signs $++-$ are $\{A',K, G, F\}$ of 8 parameters} \nn
\eea
We see that it is not necessary that the radials of the intersections of traceless matrices with various same-`dimension'-slices are accounted for by a slice-independent number of parameters. $H$ and $J$ can be of 7 parameters, whereas the radials of 
$\! \tiny \{A',K, G, F\}\!=\! \left\{ \!
\left[ \left ( \begin{array}{ccc} a  &  b   &  c \\ b & d 
& e \\ c  & e
 & -a-d \end{array} \right ) \right]\!\!\!
 :\!\!  a,d \in \mathbf{R}^{+*}\!,\! b,c,e \in \mathbf{C}
 \!\!\right\}
  $ are of 8 parameters. Moreover, we need at least 8 parameters to account for the sector COA radials ($\equiv \left(\mathcal{M}/\sim\right)^{\mbox{\tiny traceless}}$), representing the ``generalized traceless definition", since, say, the matrix $(u:u_{11}=u_{22}=u_{33}=1)$ is not equivalent to any of $\{A', F, G, K\}$, rather -through multiplication from left and right by $(\mbox{diag} \left(1,1,i\right))$-  to $(w= \in S_1: w_{11}=w_{22}=1=-w_{33})$, but it is -through multiplication from left and right by $(\mbox{diag} \left(1,\sqrt{j},j\right):j=e^{\frac{2i\pi}{3}})$- equivalent to the traceless matrix $(v: v_{11}=1, v_{22}=j, v_{33}=j^2)$.   

\begin{figure}[hbtp]
\centering
\epsfxsize=20.25cm
\centerline{\epsfbox{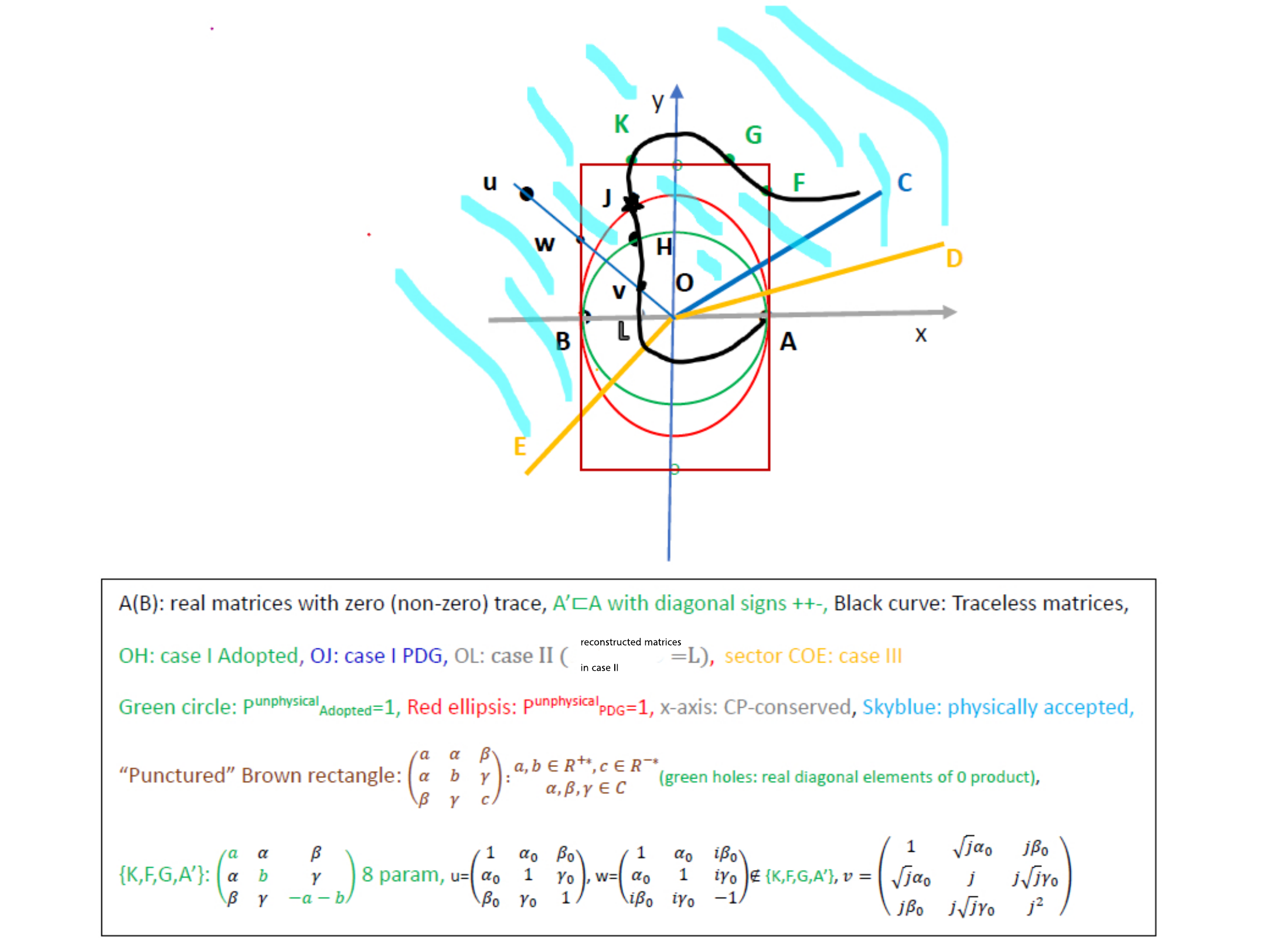}}
\caption{\footnotesize {\color{blue}Various definitions for the texture characterized by $g=0$. The  `Mathematical' def. corresponds to the submanifold defined merely by the mathematical constraint $g=0$, whereas the `Generalized' def.  (`Specific' def.)  corresponds to $g=0$ being met at one point in the radial equivalence class (at the intersection  of the radial and the vanishing unphysical slice). The number of parameters to account for the intersection of traceless matrices with a slice depends on the nature of this slice. The radials constituting the sector COA (spanned in the anticlockwise direction) are parametrised by at least 8 parameters. For any parametrization corresponding to length of base vector on $y-$axis given by $s$, we have: unphysical phases $\equiv r=\sqrt{x^2 + (y/s)^2}$, with $s=(>)1$ for Adopted (PDG), observables $\equiv \theta = \tan^{-1}(y/x)$.}} 
\label{schema}
\end{figure}

}

\bibliographystyle{unsrt}

\end{document}